\documentclass[aps,10pt,prc,tightenlines,twocolumn,twoside,showpacs,nofootinbib]{revtex4}
\usepackage{graphicx,amsmath,amssymb}
\usepackage{hyperref}
\usepackage{breakurl}

\newcommand{\eg}{\textit{e.g.}}  

\newcommand{\ie}{\textit{i.e.}}  
\newcommand{\dd}{\mathrm{d}}
\newcommand{\lsim}{\lesssim}
\newcommand{\gsim}{\gtrsim}

\begin{document}

\title{Charmonium in Medium: From Correlators to Experiment}

\author{Xingbo Zhao} \author{Ralf Rapp} \affiliation{Cyclotron Institute 
  and Physics Department, Texas A\&M University, College Station, TX
  77843-3366, USA}

\date{\today}

\begin{abstract}
 We set up a framework in which in-medium charmonium properties
 are constrained by thermal lattice QCD and subsequently implemented
 into a thermal rate equation enabling the comparison with experimental
 data in heavy-ion collisions.
 Specifically, we evaluate phenomenological consequences for charmonium 
 production originating from two different scenarios in which either the 
 free or the internal energy are identified with the in-medium 2-body 
 potential between charm and anti-charm quarks. These two scenarios
 represent $J/\psi$ ``melting temperatures" of approximately 1.25\,$T_c$
 (``weak binding") and 2\,$T_c$ (``strong binding"), respectively.
 Within current uncertainties in dissociation rates and charm-quark 
 momentum spectra, both scenarios can reproduce the centrality dependence
 of inclusive $J/\psi$ yields in nuclear collisions at SPS and RHIC
 reasonably well. However, the ``strong-binding" scenario associated
 the the internal energy as the potential tends to better reproduce 
 current data on transverse momentum spectra at both SPS and RHIC.
\end{abstract}


\maketitle

\section{Introduction}
\label{sec_intro}
Systematic studies of charmonium production in ultrarelativistic 
heavy-ion collisions (URHICs) are hoped to reveal the properties 
charm-anticharm quark bound states in hot and dense matter. A central 
goal is to utilize these insights in the search for the Quark-Gluon 
Plasma (QGP) and infer some of its basic properties, such as color
screening, flavor transport and the relevant degrees of freedom (see, 
\eg, Refs.~\cite{Rapp:2008tf,Kluberg:2009wc,BraunMunzinger:2009ih} 
for recent reviews). 
To carry out this program quantitatively requires the combination of
theoretical approaches to evaluate charmonium spectral function in medium 
with phenomenological models furnishing a realistic context for computing 
observables. The former category includes first-principle lattice-QCD (lQCD) 
calculations~\cite{Petreczky:2004pz,Aarts:2007pk,Kaczmarek:2007pb,Asakawa:2007hv} 
of charmonium correlation functions and heavy-quark (HQ) free energies, 
as well the recently revived potential 
models~\cite{Karsch:1987pv,Mocsy:2005qw,Wong:2006bx,Cabrera:2006wh,Alberico:2006vw,Mocsy:2007yj,Brambilla:2008cx,Riek:2010fk} 
to compute the charmonium spectrum in the QGP. In the latter category, 
kinetic rate equations, transport and statistical models have been
pursued~\cite{Blaizot:1996nq,Grandchamp:2003uw,Thews:2005vj,Andronic:2006ky,Yan:2006ve,Zhao:2007hh,Capella:2007jv,Linnyk:2008uf}.
However, there are currently rather few calculations with quantitative 
connections between these two categories (see, \eg, Ref.~\cite{Young:2008he}).
It is the purpose of the present paper to elaborate such connections.  

The phenomenological building block in our work consists of a kinetic 
rate equation which accounts for charmonium dissociation and 
regeneration in a thermally expanding fireball. The in-medium charmonium
properties figuring into the rate equation are inferred from spectral 
functions which are constrained by euclidean correlators and HQ free 
energies computed in thermal lQCD. The spectral functions are adopted 
from a recent potential-model calculation using a thermodynamic 
$T$-matrix~\cite{Riek:2010fk} with HQ free (or internal) energies as 
in-medium driving kernel. From these spectral functions we extract 
in-medium binding energies and bound-state masses which determine
the reaction rates and equilibrium limit for the rate equation. 
We explicitly verify that the ``reconstructed"  spectral functions
yield euclidean correlator ratios which vary little with temperature
(say, within $\pm$10\%), as found in lQCD. In view of the ongoing 
debate as to whether the free or internal energy is a more suitable
quantity to be identified with a HQ potential, we will explore both
possibilities as supplied through the $T$-matrix calculations in 
Ref.~\cite{Riek:2010fk}. As found there and in previous 
works~\cite{Wong:2006bx,Mocsy:2007yj}, the free and internal energy may 
be considered as providing an lower and upper limit, respectively, on 
the dissociation temperatures of charmonia, which in the following we 
will refer to as weak- and strong-binding scenarios. We will elaborate
the phenomenological consequences of both scenarios for charmonium
observables in URHICs at the Super Proton Synchrotron (SPS) and at 
the Relativistic Heavy-Ion Collider (RHIC), including centrality, 
transverse-momentum ($p_T$) and rapidity ($y$) dependencies. 

This paper is organized as follows: In Sec.~\ref{sec_equil} we
examine the equilibrium properties of charmonia obtained from
lQCD via the potential model and extract the key parameters needed 
for the kinetic approach of charmonium production in URHICs. In
Sec.~\ref{sec_kinetic} we review the required ingredients of our
previously constructed rate equation for calculating the
inclusive yield as well as the transverse-momentum spectra of
charmonia. In Sec.~\ref{sec_data} we compare our numerical results 
with experimental data at SPS and RHIC. We conclude and give an
outlook in Sec.~\ref{sec_concl}.

\section{Equilibrium Properties of Charmonia}
\label{sec_equil}
In this section we discuss the equilibrium properties of charmonia
($\Psi$=$J/\psi$, $\chi_c$ and $\psi'$) in the QGP medium as calculated
from lQCD and extract the temperature-dependent quantities needed in 
the kinetic approach. Specifically, these are the in-medium charm-quark 
and charmonium masses, $m^*_c$ and $m_{\Psi}$, respectively, the 
charmonium dissociation rate, $\Gamma^{\rm diss}_{\Psi}$, and the 
charmonium dissociation temperature, $T^{\Psi}_{\rm diss}$. The latter 
defines the temperature above which regeneration is not operative (\ie, 
the gain term of the rate equation is set to zero). We define
the binding energy as 
\begin{equation}
\label{epsB}
\varepsilon_B(T)=2m^*_c(T)-m_{\Psi}(T) \ .
\end{equation}  

In Sec.~\ref{ssec_corr} we review basic relations between correlators
and spectral functions and extract pertinent quantities from the 
$T$-matrix approach of Ref.~\cite{Riek:2010fk}. In Sec.~\ref{ssec_diss}
we detail our calculations of the charmonium dissociation widths which
we assume to equal the total width of the charmonium. In 
Sec.~\ref{ssec_sf} we combine the extracted quantities to 
``re-construct" charmonium spectral functions and evaluate 
corresponding euclidean correlator ratios in light of lQCD results.  

\subsection{Euclidean Correlators and Potential Model}
\label{ssec_corr}
Lattice QCD currently provides equilibrium properties of charmonium 
mainly in terms of two quantities: \\
1. The free energy, $F_{Q\bar Q}(r;T)$, of a static pair of heavy quark
and anti-quark. This is the main input for recent potential models. It 
remains controversial to date whether the free energy, the internal 
energy, 
\begin{equation}
U_{Q\bar Q}(r;T)=F_{Q\bar Q}(r;T)
-T \frac{\partial F_{Q\bar Q}(r;T)}{\partial T} \ , 
\end{equation}
or any combination 
thereof, should be identified with a static $Q\bar Q$ potential at 
finite temperature, $T$. In this work we therefore study two scenarios 
where either $U_{Q\bar Q}$ or $F_{Q\bar Q}$ is used as potential, 
$V_{Q\bar Q}$. Since the internal energy leads to 
stronger binding than the free energy, we refer to the former and 
latter as strong- and weak-binding scenario, respectively. \\
2. The two-point correlation function of a quarkonium current, 
$j_\alpha$, with hadronic quantum number $\alpha$,
\begin{equation}
G_{\alpha}(\tau,\vec r)=
\langle\langle j_\alpha(\tau,\vec r)j^\dagger_\alpha (0,\vec 0)\rangle\rangle
 \ ,
\label{Gtau}
\end{equation}
computed as a function of imaginary (euclidean) time, $\tau$ (also called 
temporal correlator). The imaginary part of the Fourier transform of the 
correlation function, $G_\alpha (\tau,\vec r)$, is commonly referred
to as the spectral function,
\begin{equation}
\sigma_\alpha (\omega,p)=-\frac{1}{\pi}\mathrm{Im}G_{\alpha}(\omega,p) \ ,
\label{spectral}
\end{equation}
which is related to the temporal correlator via
\begin{equation}
G_{\alpha}(\tau,T)
=\int^\infty_0 d\omega \sigma_\alpha (\omega,T)K(\omega,\tau,T)
\label{Gtau-sig}
\end{equation}
with the finite-$T$ kernel
\begin{equation}
K(\omega,\tau,T)=\frac{\cosh[(\omega(\tau-1/2T)]}{\sinh[\omega/2T]} \ .
\label{kernel}
\end{equation}
Lattice QCD results for two-point correlation functions are usually
normalized to a ``reconstructed" correlator evaluated with the kernel at 
temperature $T$,
\begin{equation}
G_\alpha^{\rm rec}(\tau,T)=\displaystyle\int^{\infty}_0 d\omega \ 
\sigma_\alpha(\omega,T^*) \ K(\omega,\tau,T) \ ,
\label{corr_rec}
\end{equation}
but with a spectral function at low temperature, $T^*$, where no
significant medium effects are expected. The correlator ratio,
\begin{equation}
R_\alpha(\tau,T)=G_\alpha(\tau,T)/G_\alpha^{\rm rec}(\tau,T) \ ,
\label{corr_ratio}
\end{equation}
is then an indicator of medium effects in $G_\alpha(\tau,T)$ through 
deviations from one. Current lQCD calculations find that the correlator 
ratio, $R_\alpha(\tau,T)$, in the pseudoscalar ($\eta_c$) and vector
($J/\psi$) channel are close to 1 (within ca.~10\%) at temperatures up
to 2-3\,$T_c$~\cite{Datta:2003ww,Jakovac:2006sf,Aarts:2007pk}. In the 
$P$-wave channels (scalar and axialvector) the correlators ratios are 
substantially enhanced over 1 at large $\tau$. This feature is believed 
to be due to ``zero-mode" contributions (at $\omega$=0) which are 
related to the scattering of a charm (or anti-charm) quark, $c\to c$ (or 
$\bar c\to \bar c$), rather than to $c\bar c$ bound-state 
properties~\cite{Umeda:2007hy}. This interpretation is supported by
studies of the $\tau$-derivative of $P$-wave correlator ratios, which 
exhibits a much smaller variation (in the limit that the zero-mode part 
is a $\delta$-function, $\sigma_{\rm zm}(\omega)\propto\delta(\omega)$, 
its contribution to the temporal correlator is a 
constant)~\cite{Umeda:2007gk,Petreczky:2008px}. 

In principle, the in-medium properties of charmonia, such as pole mass, 
in-medium width and dissociation temperature, are fully encoded in their 
spectral function. However, the finite number of data points for the 
two-point correlator computed in lQCD severely hampers the inversion of 
the transform in Eq.~(\ref{Gtau-sig}), rendering the determination of 
the spectral function difficult. In fact, potential-model analyses have 
shown that the use of either the free {\em or} internal energy can lead 
to agreement with lQCD correlators, albeit with rather different 
underlying binding properties (and associated dissociation temperatures). 
In the present work we therefore adopt the following strategy: we first 
extract the charm-quark masses and charmonium pole mass from a potential 
model, allowing us to define the 
binding energy according to Eq.~(\ref{epsB}). The latter is an important 
ingredient in the quantitative evaluation of the $\Psi$ dissociation 
rate~\cite{Grandchamp:2001pf}, which will be done in the following
section (\ref{ssec_diss}) in a perturbative (``quasifree")
approximation. We then ``reconstruct" in-medium charmonium spectral
functions using a relativistic Breit-Wigner + continuum ansatz, where 
the $\Psi$ width and mass figure into the Breit-Wigner part while the 
continuum is determined by the open-charm threshold (2$m_c^*$). For a 
more realistic evaluation, we include polestrength factor, $Z_\Psi(T)$, 
for the Breit-Wigner strength and a nonperturbative rescattering 
enhancement in the continuum~\cite{Cabrera:2006wh,Mocsy:2007yj}.
The vanishing of the polestrength factor furthermore serves to estimate
the dissociation temperature of the ground state in each channel.

For definiteness we employ the potential model of Ref.~\cite{Riek:2010fk}
where quarkonium spectral functions and correlators have been calculated 
in a thermodynamic $T$-matrix approach, consistent with vacuum spectroscopy
and including relativistic corrections for a proper description of 
scattering states. The calculations in there have been carried out
for both free and internal energies as potential, and for two different
lQCD inputs~\cite{Petreczky:2004pz,Kaczmarek:2007pb}. In both cases (and 
for both potentials), an approximate
constancy (within $\pm$15\%) of the correlator ratios for pseudoscalar
charmonium has been found (see lower panels of Fig.~12 and 14 in 
Ref.~\cite{Riek:2010fk}). We believe that these results provide a
reasonable representation and bracket for potential-model results.  
\begin{figure}[!t]
\centering
\includegraphics[angle=-90,width=0.48\textwidth]{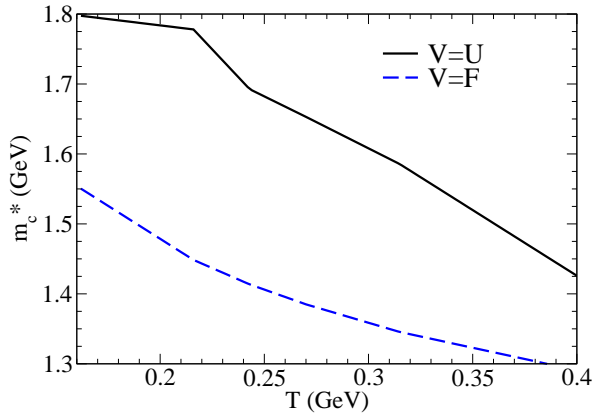}
\caption{(Color online) Temperature dependence of in-medium charm
  quark mass. The solid (dashed) line is for the strong (weak) binding
  scenario. }
\label{fig_mc}
\end{figure}
In Fig.~\ref{fig_mc} the temperature-dependent charm-quark mass is 
displayed, which is identified with the asymptotic value of 
the HQ potential,
\begin{equation}
m^*_c(T)\equiv m^0_c+V_{Q\bar Q}(r\to\infty;T)/2 \ . 
\end{equation}
The in-medium masses decrease with temperature appreciably, while the 
magnitude of $m^*_c(T)$ is significantly smaller in the
weak-binding compared to the strong-binding scenario. 
As expected, the binding energies (plotted in Fig.~\ref{fig_epsB}) 
also decrease with $T$, again being significantly smaller in the
weak-binding scenario. These features are, in fact, the main reason 
that both scenarios can be compatible with 
the small variations found in the lQCD correlator ratios: for 
weak/strong binding, a small/large constituent mass combines 
with a small/large binding energy, respectively, leading to an 
approximate compensation in the bound-state mass,
$m_\Psi(T)$, recall Eq.(\ref{epsB}).
\begin{figure}[!t]
\centering
\includegraphics[angle=-90,width=0.48\textwidth]{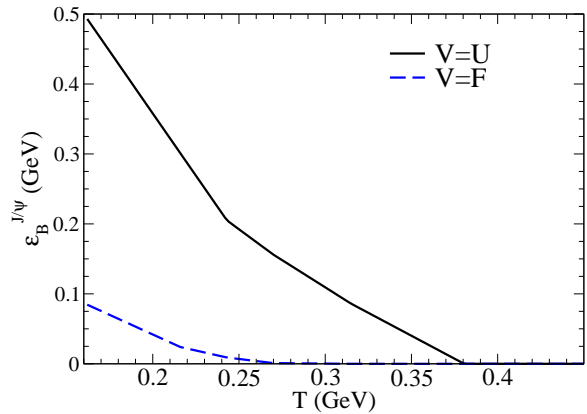}
\caption{(Color online) Temperature dependence of $J/\psi$ binding
  energy. The solid (dashed) line is for the strong (weak) binding
  scenario with $V_{Q\bar Q}$=$U_{Q\bar Q}$($F_{Q\bar Q}$). }
\label{fig_epsB}
\end{figure}

\subsection{Dissociation Rates}
\label{ssec_diss}
The inelastic dissociation rate of charmonia in the medium, 
$\Gamma^{\rm diss}_{\Psi}(T)$, plays a central role in URHIC 
phenomenology as it directly governs the time dependence of
their abundance in an underlying rate equation (see, \eg, 
Eq.~(\ref{rate-eq}) below). It contributes to the total width of 
the corresponding charmonium spectral function (in addition to 
elastic scattering which we neglect in the present work). 
However, it is currently not possible to quantitatively extract the
width from temporal correlators~\cite{Umeda:2005pk}: for 
phenomenologically relevant values of the width of a few tens
of MeV (or even up to 200\,MeV) the correlator ratios are affected 
at a level of $\sim$5\%~\cite{Cabrera:2006wh,Riek:2010fk} which is too 
small to be discerned from other uncertainties at this point.   
Furthermore, elastic and inelastic collisions contribute to the width
with an a priori unknown partition (for relatively loosely bound states 
one expects the dissociation width to be dominant). In the present work 
we therefore calculate the dissociation rate based on perturbative QCD 
with an effective strong coupling constant which we later fine tune to 
reproduce the observed $J/\psi$ suppression in central A-A collisions. 
The resulting value of $\alpha_s$ turns out to be $\sim$0.3, quite 
compatible with the short-distance (color-Coulomb) term in the effective 
potential used to extract the binding energies discussed in the previous 
section.

Let us briefly discuss basic mechanisms for charmonium 
dissociation in medium. In the QGP, the leading-order process is naively 
given by the gluo-dissociation process introduced more than 30 years 
ago~\cite{Peskin:1979va}, $\Psi+g\to c+\bar c$.\footnote{In the language 
of effective field theory (EFT)~\cite{Brambilla:2008cx}, this corresponds to 
the color-singlet to -octet transition, albeit final-state interactions
in the $c+\bar c$ octet state are neglected here. However, they are
repulsive and of order ${\cal O}(1/N_c^2)$ which renders them numerically 
very small.} However, for small binding energies this process
becomes inefficient (due to a shrinking phase space) and is superseded 
by inelastic reactions with an extra parton in the final state,  
$i + \Psi \to i+c+\bar c$ 
($i$=$g$,$q$,$\bar q$)~\cite{Grandchamp:2001pf}.\footnote{In EFT
language, some of these processes (\eg, the ones involving $t$-channel
gluon exchange between the thermal parton $i$ and one of the
charm quarks in the bound state) correspond to the Landau damping
contribution to the width.}   
Following Ref.~\cite{Grandchamp:2001pf}, we treat this process
as a ``quasi-elastic" collision between the parton $i$ from the medium 
and the $c$ or $\bar c$ quark in the bound state,  $i + c'
\to i+c$. In this quasifree approximation the charmonium dissociation 
cross section becomes twice the elastic cross section between the parton
and the $c$ quark,  $\sigma_\Psi^{\rm diss}=2\sigma^{\rm el}_{ci}$,
as, \eg, given in Ref.~\cite{Combridge:1978kx}. To account for the 
leading kinematic correction from the residual binding energy, the 
incoming parton needs to be energetic enough to break up the bound 
state, which sets a lower limit for the incoming parton momentum.
Overall 4-momentum conservation for the process $i+\Psi \to i+c+\bar c$ 
is maintained by assigning the binding energy
to a decrease in mass of the initial-state charm-quark, $c'$, \ie,
$m_{c'} = m_c -\varepsilon_B$. In addition, we have introduced a Debye 
mass, $m_D=gT$, into the denominator of $t$-channel gluon-exchange 
propagator, $1/t\to 1/(t-m^2_D)$, to regulate the divergence for
forward scattering (the strong coupling in $m_D$ is taken consistently
with the coupling constant $\alpha_s$). 

In the hadron gas (HG) phase, we employ a flavor-$SU(4)$ effective 
Lagrangian approach~\cite{Lin:1999ad,Haglin:2000ar} to estimate the
inelastic cross sections with $\pi$ and $\rho$ mesons. As we will see
below most of the charmonium dissociation in the hot medium occurs in
the QGP, because the typical density of hadrons in HG is much smaller 
than that of partons in QGP, while the dissociation cross sections are 
of similar magnitude (around 1\,mb).

\begin{figure}[!t]
\centering
\includegraphics[angle=-90,width=0.48\textwidth]{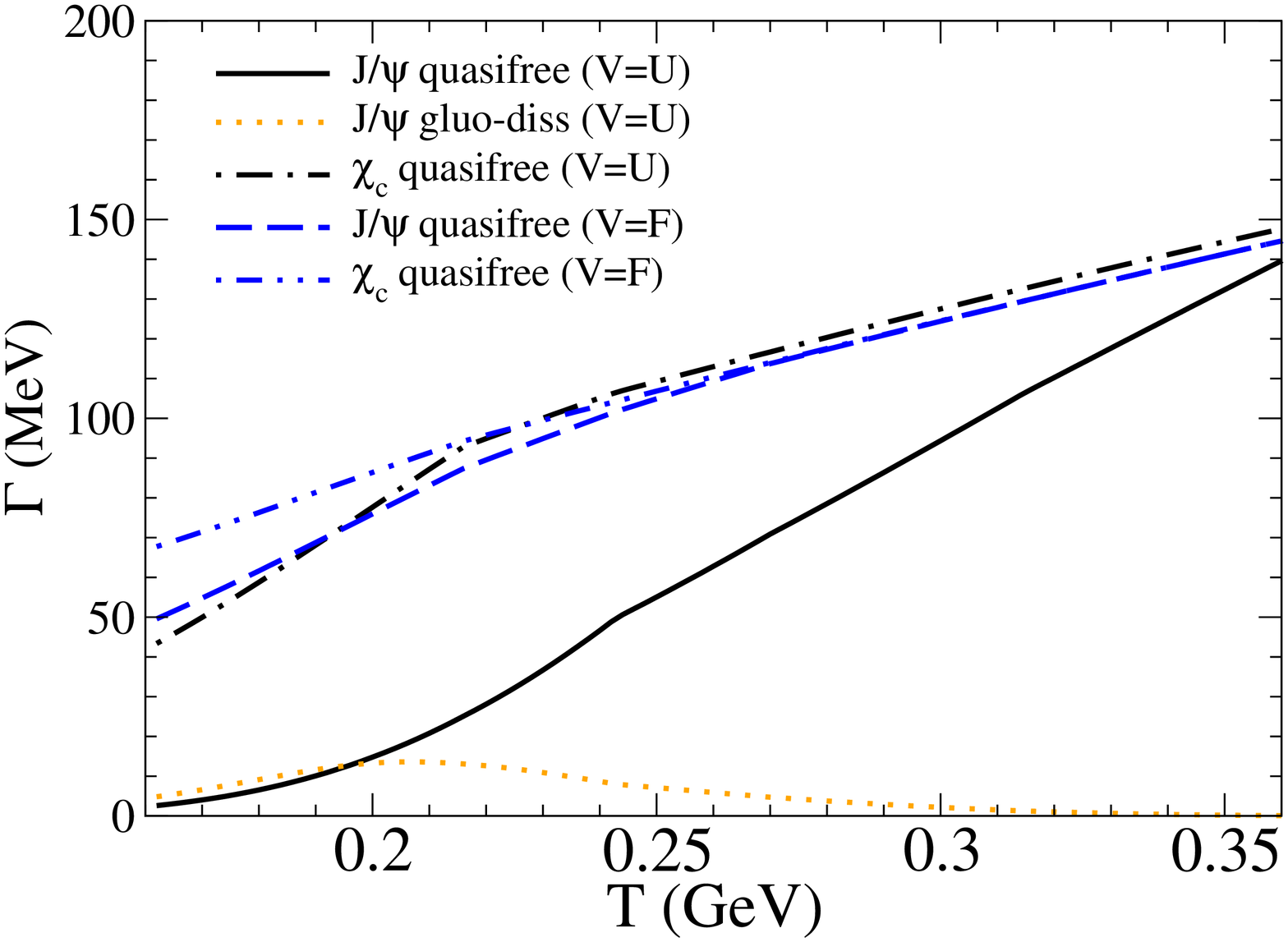}

\vspace{-0.8cm}

\includegraphics[angle=-90,width=0.48\textwidth]{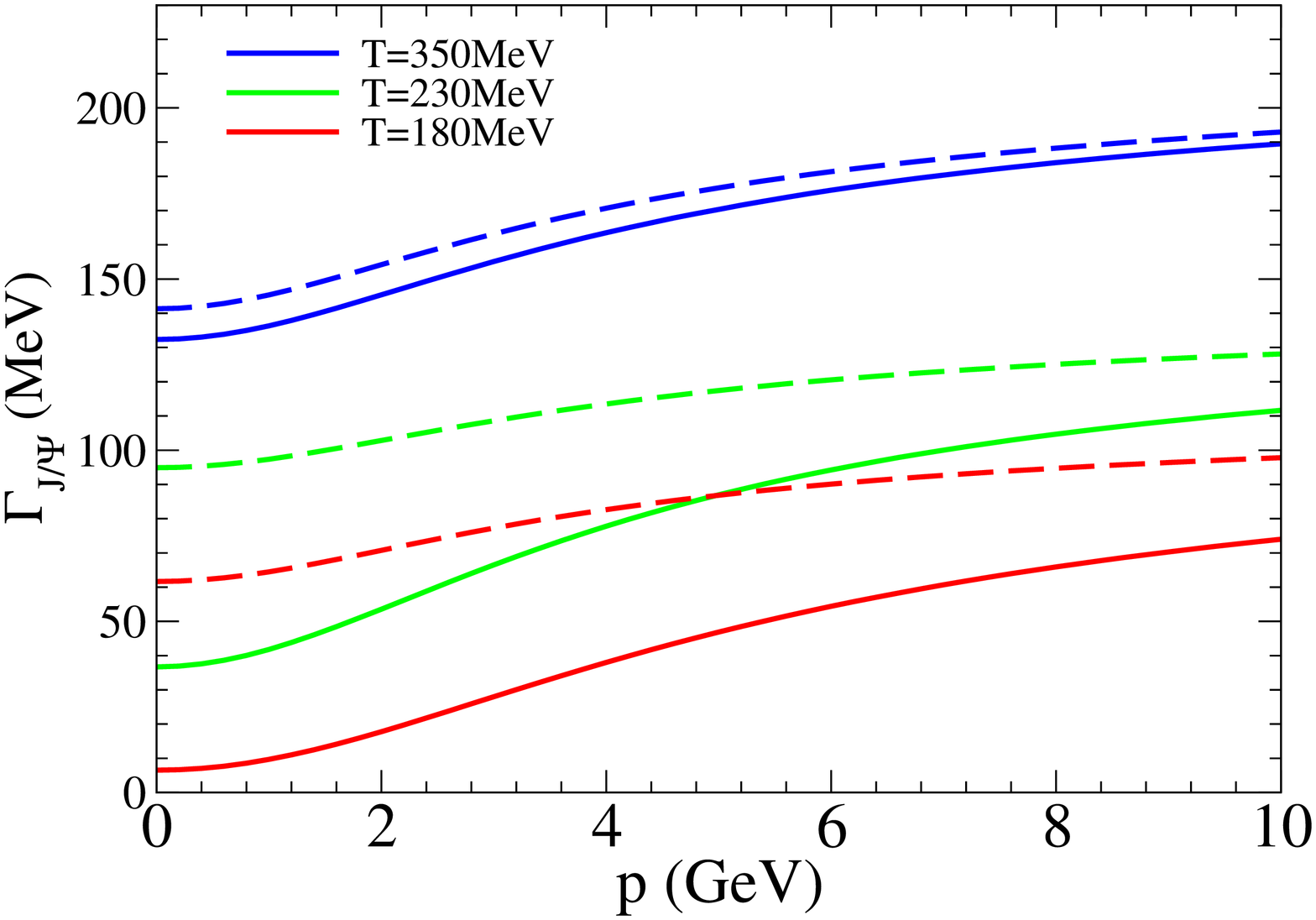}
\caption{(Color online) Upper panel: temperature dependence of 
 dissociation rates for $J/\psi$ and $\chi_c$ calculated in quasifree 
 approximation for the strong-binding (solid line: $J/\psi$, dot-dashed 
 line: $\chi_c$) and weak-binding scenarios (dashed line: $J/\psi$, 
 dash-double-dotted line: $\chi_c$). The dotted line indicates the 
 contribution from $J/\psi$ gluo-dissociation rate in the strong binding 
 scenario using $\alpha_s$=0.32 in the underlying cross section 
 expression~\protect\cite{Peskin:1979va}.
 Lower panel: 3-momentum dependence of the $J/\psi$ dissociation rate
 in the strong- and weak-binding scenarios (solid and dashed lines,
 respectively).}
\label{fig_diss-rate}
\end{figure}

The dissociation rate of a charmonium state at finite 3-momentum, $p$,
can be obtained from the inelastic cross section by a convolution over 
the thermal distribution, $f^i(\omega;T)$,
of medium particles in the QGP or HG,
\begin{equation}
\Gamma_\Psi^{\rm diss}(p,T) 
= \sum_{i} \int \frac{d^3k}{(2\pi)^3} \ f^i(\omega_k;T) 
 \  \sigma^{\rm diss}_{\Psi i}(\vec p,\vec k)\ v_{\rm rel} \ .
\label{rate}
\end{equation}
Here, $v_{\rm rel}$ is the relative velocity of $\Psi$ and medium particle
$i$. The temperature dependence of the quasifree dissociation rates 
for the $J/\psi$ and $\chi_c$ in the QGP are plotted at $\vec p$=0
vs. $T$ in the upper panel of Fig.~\ref{fig_diss-rate}, and for the
$J/\psi$ as a function of $p$ at selected temperatures in the lower panel.
In the weak-binding scenario, there is rather little difference
between the dissociation rates of $J/\psi$ and $\chi_c$, especially
above $T=200$\,MeV. Only in the strong-binding scenario the larger
$J/\psi$ binding energy makes a large difference, suppressing its
destruction by, \eg, a factor of $\sim$5 at $T\simeq$200\,MeV relative
to the $\chi_c$ and $\psi'$ (not shown); this difference becomes
larger (smaller) at smaller (larger) $T$. For comparison we also
calculated the rate due to the gluo-dissociation mechanism employing
the expression derived in Ref.~\cite{Peskin:1979va} with the same
$\alpha_s$=0.32 as in the quasifree rate and with $\epsilon_B$
obtained from the strong-binding scenario (note that the Coulombic
binding is much smaller), which turns out to be inefficient for
dissociating $J/\psi$'s (and even more so for the excited states) and
is thus neglected in the following. The 3-momentum dependence of
the rates shows a monotonous increase with increasing $p$, which
becomes more pronounced with increasing binding energy (for larger
$\varepsilon_B$ a finite 3-momentum facilitates the break-up since, on
average, a larger center-of-mass energy is available in the collision
of the bound state with thermal partons). This increase is a simple
kinematic consequence of a monotonously increasing (or even constant)
cross section with finite threshold and an increasing parton flux
encountered by the moving $J/\psi$.

In the following section we determine the interplay of the charmonium 
widths with their binding energies extracted in the previous section
in order to assess their survival in the QGP (\eg, as a resonance).

\subsection{Spectral Functions}
\label{ssec_sf}
We now turn to the construction of the charmonium spectral functions 
and the corresponding two-point correlation functions, and constrain 
the latter by lQCD computations. In doing so we introduce an 
additional quantity, namely the polestrength, $Z_\Psi(T)$, of the 
Breit-Wigner bound-state part characterizing the disappearance of 
the state from the spectrum. The vanishing of $Z_\Psi(T)$ will be 
used to characterize the dissolution temperature above which 
regeneration in the rate equation is inoperative.

We first construct a model spectral function in vacuum, consisting of a
zero-width bound-state and a perturbative (leading order) continuum part,
\begin{multline}
\sigma_\Psi (\omega)=A_\Psi \ \delta(\omega-m_{\Psi})\\
+\frac{B_\Psi N_c}{8\pi^2}\Theta(\omega-s_0)\omega^2
\sqrt{1-\frac{s^2_0}{\omega^2}}(a+b\frac{s^2_0}{\omega^2}) \ .
\label{spf}
\end{multline}
Here, $N_c$=3 is the number of colors and the coefficients 
$(a,b)=(1,-1), (2,1)$ characterize the scalar and vector channel, 
respectively~\cite{Mocsy:2005qw}. The open-charm threshold in vacuum, 
$s_0$, is assumed to be given by twice the free $D$- meson mass,
$s_0\equiv 2m_D=3.74$\,GeV. The coefficient $A$ is related to the 
overlap of the wave-function, $R_{J/\psi}(0)$, or its derivative, 
$R'_{\chi_{c}}(0)$, at the origin~\cite{Bodwin:1994jh,Mocsy:2005qw},
\begin{equation}
  A_{J/\psi}=\frac{3N_c}{2\pi}|R_{J/\psi}(0)|^2 \ , \ \ 
  A_{\chi_{c}}=\frac{36N_c}{2\pi M^2_{\chi_{c}}}|R'_{\chi_c}(0)|^2 \ .
\label{norm_res}
\end{equation}
These quantities can be estimated from 
the electromagnetic decays widths via~\cite{Bodwin:1994jh}
\begin{equation}
\Gamma_{ee}=\frac{4e_Q^2\alpha^2N_c}{3m^2_{J/\psi}}|R_{J/\psi}(0)|^2 
\ , \ 
\Gamma_{\gamma\gamma}=\frac{144e_Q^4\alpha^2N_c}{m^4_{\chi_{c}}}
|R'_{\chi_c}(0)|^2
\label{gamma_em}
\end{equation}
where $\alpha$=1/137 is the electromagnetic coupling constant and
$e_Q=2/3$ the charge of the charm quark (we use 
$\Gamma_{ee}$=5.55\,keV for the $J/\psi$ and 
$\Gamma_{\gamma\gamma}$=2.40\,keV for the $\chi_{c0}$). The resulting 
relations between $A_\Psi$ and $\Gamma_{\Psi\to ee,\gamma\gamma}$
are
\begin{equation}
A_{J/\psi}=\frac{81m^2_{J/\psi}}{32\pi\alpha^2} \ \Gamma_{ee} \ , \ \ 
A_{\chi_{c}} =\frac{81m^2_{\chi_{c0}}}{128\pi\alpha^2}  
\ \Gamma_{\gamma\gamma} \ .
\label{norm_r}
\end{equation}
The $J/\psi$ and $\chi_{c0}$ masses are taken at their empirical 
vacuum values.
The coefficient $B_\Psi$ in the continuum part of Eq.~(\ref{spf}) equals
one in the non-interacting limit. To account for rescattering, which
is particularly important close to threshold, we scale it up to
match the continuum as calculated from the vacuum $T$-matrix in 
Ref.~\cite{Riek:2010fk}, amounting to $B_{J/\Psi}\simeq2$ and 
$B_{\chi_{c}}\simeq4$ in the vector and scalar channel, respectively.
For simplicity we neglect $\psi'$, $\chi_{c}'$ and higher excited 
states which play little role in the correlator ratios.

At finite temperature we replace the $\delta$-function bound-state
part by a relativistic Breit-Wigner (RBW) distribution while the
continuum part is assumed to be of the same form as in the vacuum,
\begin{align}
\label{spfT} 
\sigma_\Psi(\omega)=A_\Psi \ Z_\Psi(T) \frac{2\omega}{\pi}
\frac{\omega\Gamma_\Psi(T)}{(\omega^2-m_{\Psi}^2(T))^2+\omega^2
\Gamma_\Psi(T)^2} \qquad
\nonumber\\
\ +\frac{B_\Psi N_c}{8\pi^2}\Theta(\omega-s(T))\omega^2 
\sqrt{1-\frac{s(T)^2}{\omega^2}}(a+b\frac{s(T)^2}{\omega^2}) \ .
\nonumber\\
\end{align}
The in-medium continuum edge, $s(T)$, is now taken as the charm-quark
threshold at finite temperature, $s(T)\equiv2m^*_c(T)$, consistent with
the potential model, see Fig.~\ref{fig_mc}. The RBW term includes the 
in-medium charmonium mass, $m_\Psi(T)$, extracted from Eq.~(\ref{epsB})
based on Figs.~\ref{fig_mc} and \ref{fig_epsB}, the width $\Gamma_\Psi$
identified with the inelastic dissociation width discussed in the
previous section, and the aforementioned polestrength factor, 
$Z_\Psi(T)$. The latter is adjusted to minimize the deviation of 
the correlator ratios from one.
\begin{figure}[!t]
\centering
\includegraphics[angle=-90,width=0.48\textwidth]{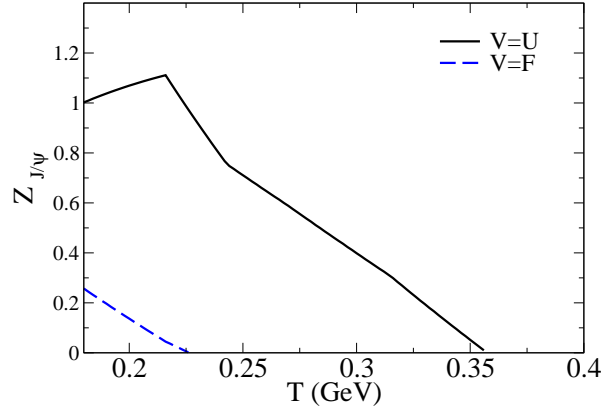}
\caption{(Color online) Temperature dependence of the strength of the
  resonance part of the $S$-wave spectral function, $Z_\Psi(T)$. The 
  solid (dashed) line is for the strong (weak) binding scenario.}
\label{fig_zpsi}
\end{figure}
The resulting $Z(T)$ for $J/\psi$ (vector channel) is plotted in 
Fig.~\ref{fig_zpsi}, from which we extract its dissociation temperature
$T_{J/\psi}^{\rm diss}$=2.0(1.25)$T_c$ in the strong (weak) binding 
scenario. Similar analysis in the scalar channel yields $\chi_{c}$ 
dissociation temperatures of  $T_{\chi_{c}}^{\rm diss}$=1.3(1.0)$T_c$ 
in the strong (weak) binding scenarios. We assume that $\chi_{c1}$ and
$\chi_{c2}$ have the same dissociation temperatures as the
$\chi_{c0}$. For $\psi'$ we simply assume its dissociation temperature
to be $T_c$ for both the strong- and weak-binding scenarios.

\begin{figure}[!t]
\includegraphics[angle=-90,width=0.48\textwidth]{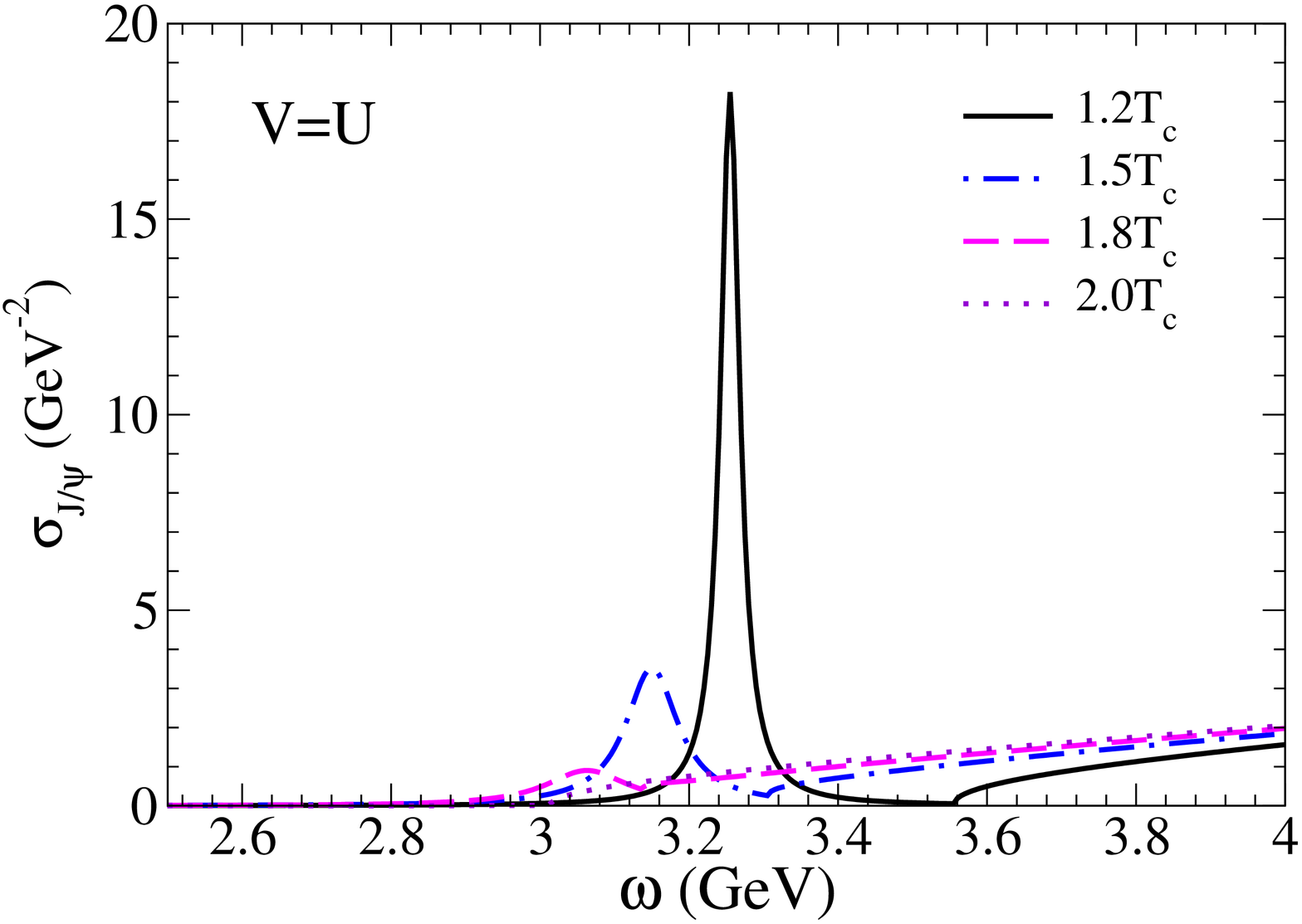}

\vspace{-0.8cm}

\includegraphics[angle=-90,width=0.48\textwidth]{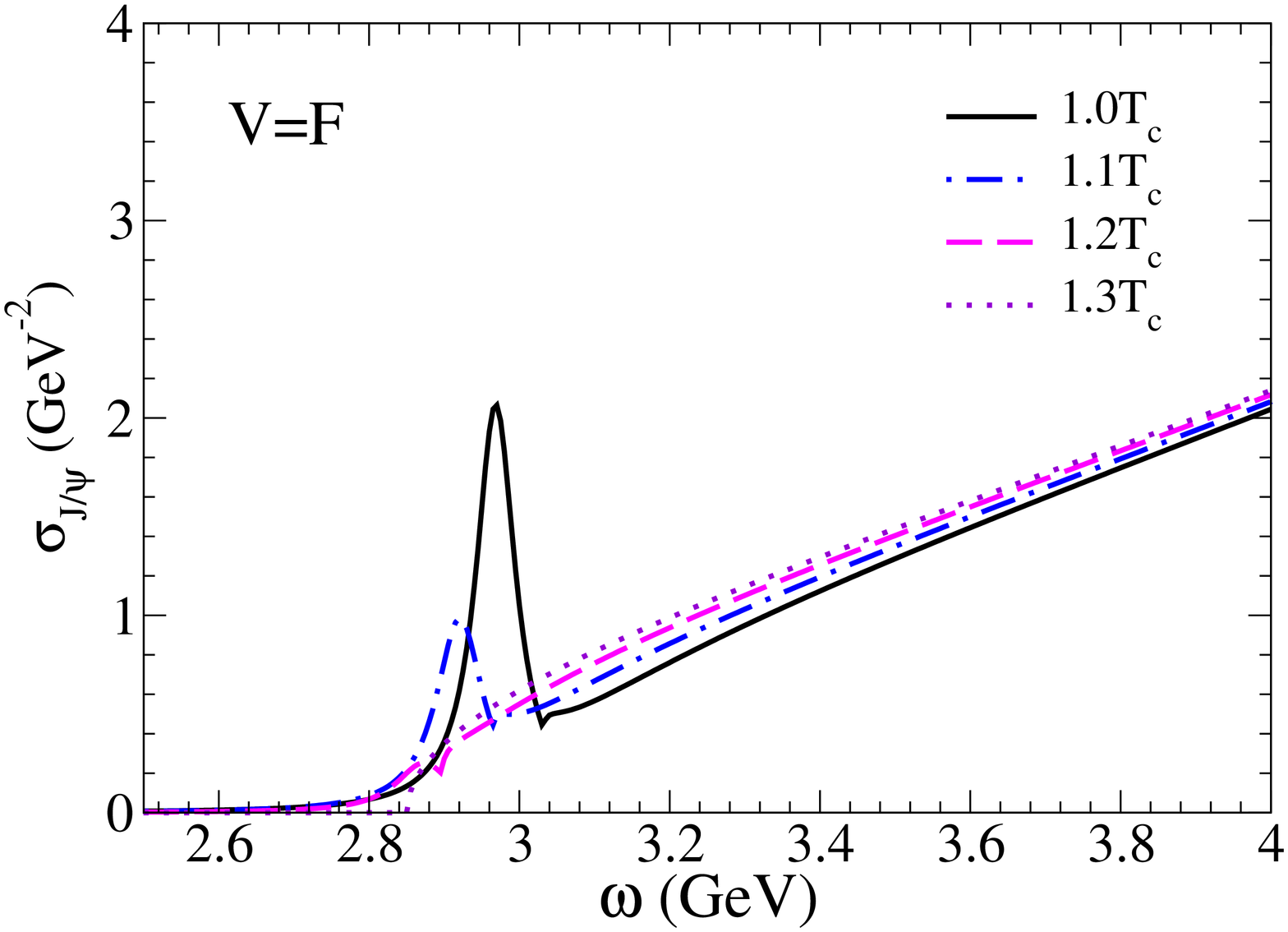}
\caption{(Color online) Spectral functions in the vector channel. The upper (lower) panel is for the strong (weak) binding
  scenario. }
\label{fig_sf-jpsi}
\end{figure}

\begin{figure}[!t]
\centering
\includegraphics[angle=-90,width=0.48\textwidth]
{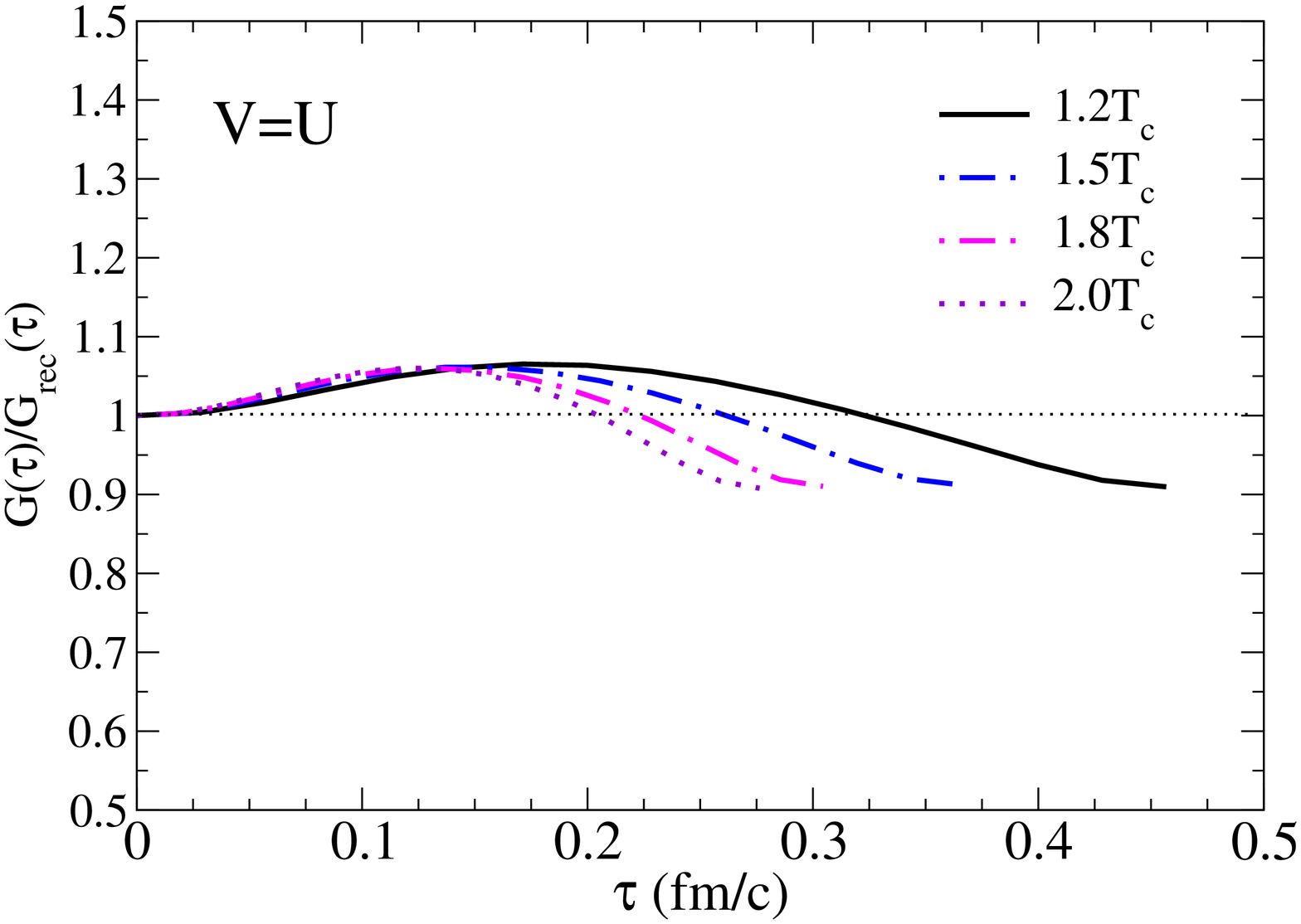}

\vspace{-0.8cm}  

\includegraphics[angle=-90,width=0.48\textwidth]
{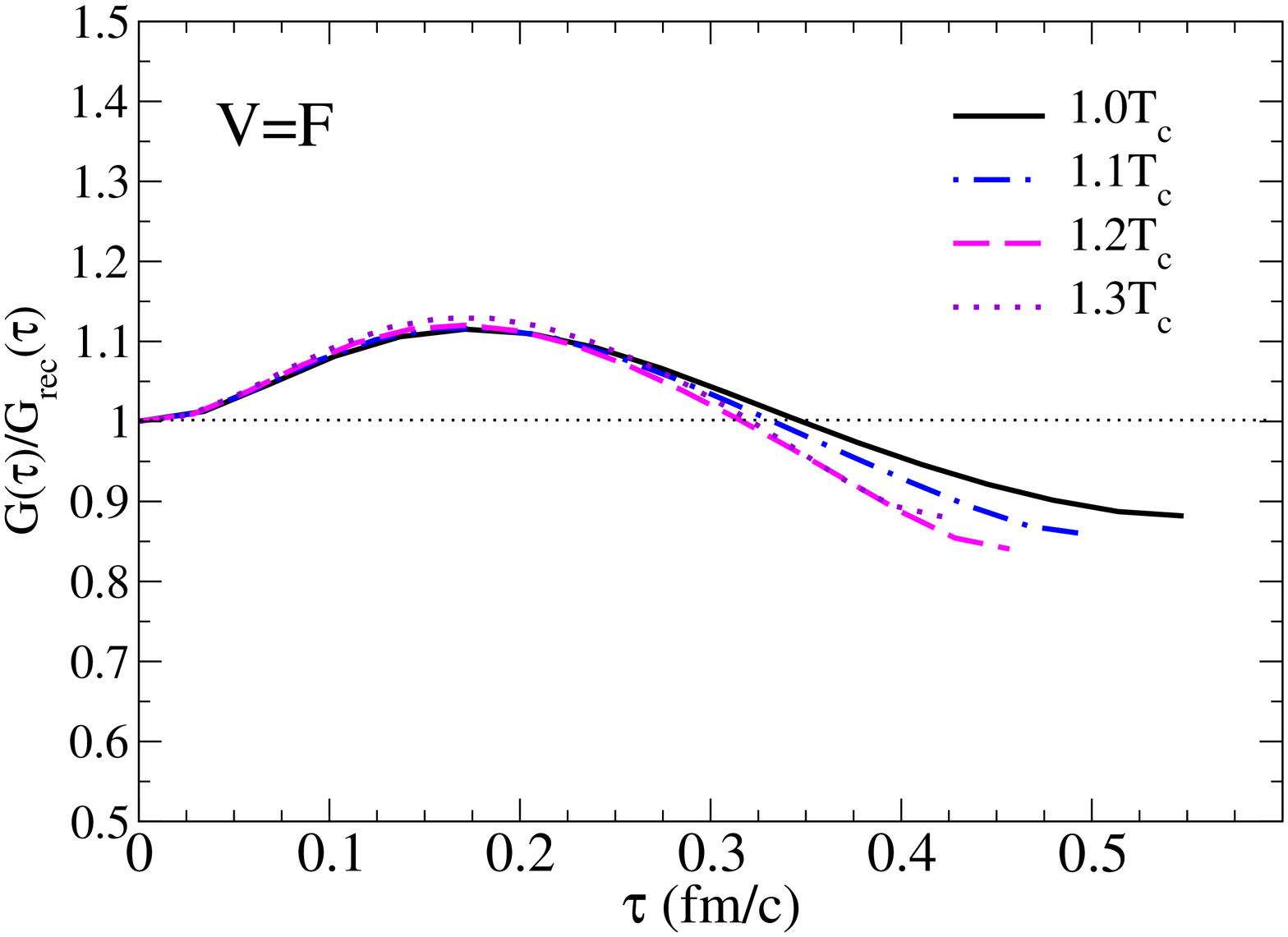}
\caption{(Color online) Ratio of vector channel correlator to the
  reconstructed correlator. The upper (lower) panel is for the strong
  (weak) binding scenario. }
\label{fig_corr-jpsi}
\end{figure}

\begin{figure}[!t]
\centering
\includegraphics[angle=-90,width=0.48\textwidth]{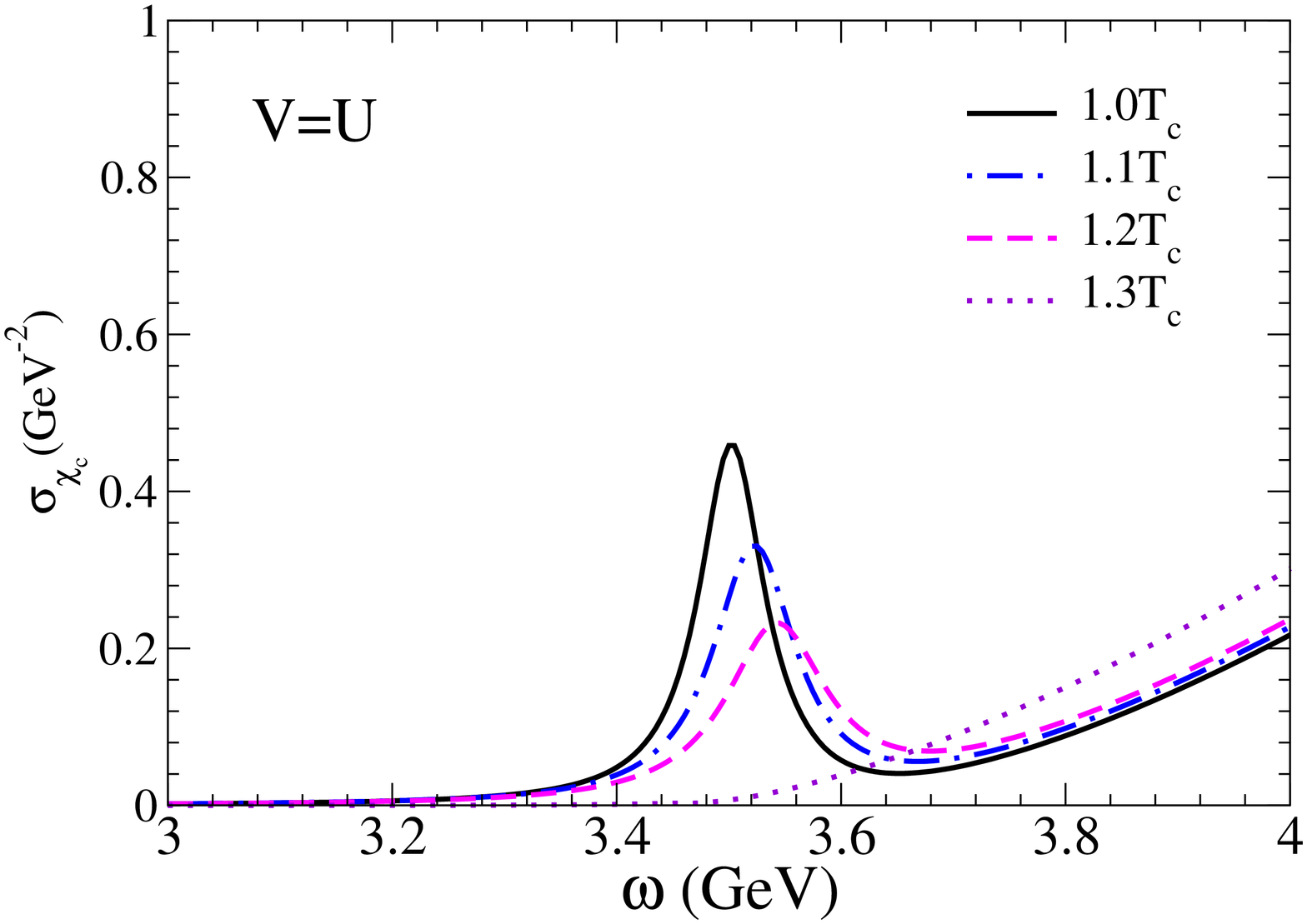}

\vspace{-0.8cm}

\includegraphics[angle=-90,width=0.48\textwidth]{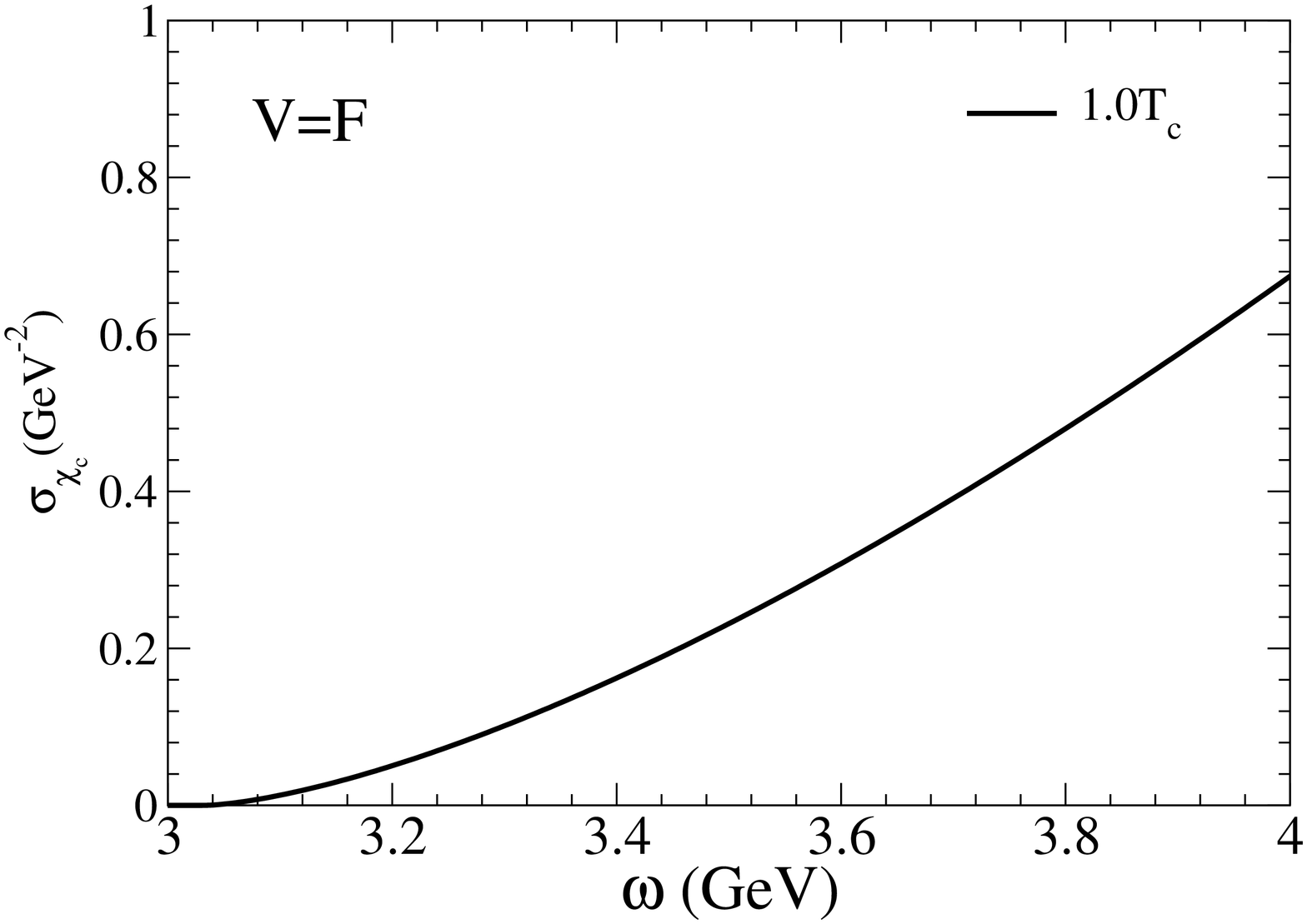}
\caption{(Color online) Spectral functions in the scalar channel. The
  upper (lower) panel is for the strong (weak) binding scenario. In
  the weak-binding scenario $\chi_{c0}$ has already melted at $T_c$.}
\label{fig_sf-chi}
\end{figure}

To comprehensively illustrate the medium effects we plot the final 
spectral functions for the vector channel in the strong- and 
weak-binding 
scenario in the QGP in Fig.~\ref{fig_sf-jpsi}, and their corresponding 
correlator ratios in Fig.~\ref{fig_corr-jpsi}; the spectral functions for the
scalar channel are displayed in Fig.~\ref{fig_sf-chi}. 
We see that the correlator ratios are indeed close to one, as
found in lQCD. In the hadronic phase (not shown), we assume vacuum
masses for both charmonia and open-charm hadrons, which automatically
ensures that the correlator ratios are close to one (deviations
due to small charmonium widths in hadronic matter are negligible).

We are now in position to implement the in-medium properties of
the charmonia into a kinetic rate equation in a thermal background,
paving the way for applications to experimental data in heavy-ion
collisions.  

\section{Kinetic Approach}
\label{sec_kinetic}
The evolution of charmonium yields and spectra in a nucleus-nucleus
($A$-$A$) collision can be roughly divided into three stages. In the 
first, ``primordial" stage (at small times $\tau\simeq0$) charm-quark 
pairs are produced in initial hard nucleon-nucleon ($N$-$N$) collisions. 
In the second, ``pre-equilibrium" stage ($0\lsim\tau\lsim\tau_0$) 
charmonia are typically in the formation phase, but the 
so-called pre-resonance states are already subject to 
dissociation by passing-by nucleons. The third, ``equilibrium" stage 
($\tau\gsim\tau_0$) starts once the hot and dense medium has 
thermalized and lasts until thermal freezeout after which hadrons
stream freely to the detectors. During this stage charmonia are 
subject to dissociation by particles of the heatbath, but detailed
balance requires that $c\bar c$ pairs in the medium can also recombine 
and ``regenerate" charmonia (below their respective dissociation 
temperature). 

Guided by the above scheme, we have organized the remainder of this 
section as follows: In Sec.~\ref{ssec_cnm} 
we discuss primordial charmonium production and its modifications in 
the pre-equilibrium stage, including rapidity and transverse-momentum 
dependencies. In Sec.~\ref{ssec_fb} we briefly recapitulate a simple 
fireball model which serves as a thermally evolving background for the 
charmonium rate equation. The latter is introduced in 
Sec.~\ref{ssec_rate} along with its main ingredients related to 
in-medium open-charm and charmonium properties. In 
Sec.~\ref{ssec_pt-dep} we specifically address the transverse-momentum 
dependence of charmonium production in the equilibrium stage.

\subsection{Cold Nuclear Matter Effects}
\label{ssec_cnm}
Charmonium production yields and $p_t$ distributions following from 
the primordial and pre-equilibrium stage form the initial conditions 
for the thermal rate equation discussed in the next section.  
The starting point are primordial spectra based on experimental data 
in $p$-$p$ collisions, modified by corrections specific to proton-nucleus
($p$-$A$) and extrapolated to $A$-$A$ collisions. The corrections are
commonly referred to as cold-nuclear-matter (CNM) effects, which include 
the following: \\
1) Nuclear shadowing, the modification of the initial parton distribution 
functions in a nucleus relative to those in a proton, which affects 
both open and hidden charm yields in $A$-$A$ relative to $p$-$p$ 
collisions.  \\
2) Cronin effect, the increase of the mean $p_t$ of produced charmonia 
in $A$-$A$ relative to $p$-$p$ associated with initial-state parton 
scattering prior to the hard production.  \\
3) Nuclear absorption, the dissociation of pre-charmonium states 
by passing-by nucleons.

In the present work, we assume a factorization of the charmonia 
phase-space distribution function into spatial and momentum parts,
 \begin{eqnarray}
 f_\Psi(b,\vec x_t,\vec p_t,\tau_0)= 
f_\Psi(b,\vec x_t,\tau_0) \ f_\Psi(b,\vec p_t,\tau_0)
\label{CNM}
\end{eqnarray} 
($b$: impact parameter of the $A$-$A$ collision, which we use
as a measure of centrality equivalent to $N_{\rm part}$, the
number of nucleon participants). 
The Cronin effect is readily implemented into the 3-momentum dependent 
part $f_\Psi(b,\vec{p_t},\tau_0)$ via a Gaussian smearing of the 
charmonium $p_t$ distribution in $p$-$p$ collisions, $f_\Psi^{pp}(p_t)$,
\begin{equation}
  f_\Psi(b,\vec{p_t},\tau_0)=\int\frac{d^2q_t}{2\pi\langle\Delta p_t^2\rangle}
\exp{\left(-\frac{q_t^2}{2 \langle \Delta p_t^2\rangle}\right )}\
  f_\Psi^{pp}(|\vec{p}_t-\vec{q}_t|) \ .  
\label{Cronin}
\end{equation}
The nuclear increase of the average $p_t^2$, 
$\langle \Delta p_t^2\rangle = 
\langle p_t^2\rangle_{AA}-\langle p_t^2\rangle_{pp}$,
is estimated within a random-walk treatment of parton-nucleon 
collisions~\cite{Huefner:2002tt} as being proportional to the mean 
parton path length, $\langle l \rangle$, in the cold medium: 
$ \langle \Delta p^2_t\rangle=a_{gN}~\langle l\rangle$. The
coefficient $a_{gN}$ is estimated from $p$-$A$ data at
SPS~\cite{Topilskaya:2003iy} and d-Au data at
RHIC~\cite{Adare:2007gn}. We use $a_{gN}$=0.076\,GeV$^2$/fm
for $\sqrt s$=17.3\,AGeV Pb-Pb collisions and
$a_{gN}$=0.1(0.2)\,GeV$^2$/fm for $\sqrt s$=200\,AGeV Au-Au collisions 
at mid (forward) rapidity. 

Nuclear absorption is evaluated within a Glauber model; the resulting
charmonium distribution in the transverse plane takes the form
\begin{align}
  f_\Psi(\vec{b},\vec{x_t},\tau_0)=\Delta y\frac{\dd\sigma^\Psi_{pp}}{dy}
  \int dz\ dz^{\prime}\rho_A(\vec{x}_t,z)\
  \rho_B(\vec{b}-\vec{x}_t,z^{\prime})
  \nonumber\\
  \times\exp\left\{-\int\limits^\infty_z
    dz_A\rho_A(\vec{x}_t,z_A)\sigma_{\rm abs}\right\} \qquad
  \nonumber\\
  \times\exp\left\{-\int\limits^\infty_{z^\prime}
    dz_B\rho_B(\vec{b}-\vec{x}_t,z_B)\sigma_{\rm abs}\right\}  
\label{glauber}
\end{align}
where $\rho_{A,B}$ are Woods-Saxon profiles~\cite{De Jager:1974dg} of nuclei
$A$ and $B$ and $\Delta y$=1.8 represents the rapidity coverage of
our thermal fireball (see Sec.~\ref{ssec_fb} below). For the $pp$ charmonium production cross per unit rapidity
we take the values $\dd \sigma_{pp}^{\Psi}/dy$=37\,nb~\cite{Abt:2005qr}
for $\sqrt s$=17.3\,AGeV Pb-Pb~\cite{Abt:2005qr} (with ca.~40\%
uncertainty) and $\dd \sigma_{pp}^{\Psi}/dy$=750(500)\,nb for 
$\sqrt s$=200\,AGeV Au-Au~\cite{Adare:2006kf} at mid and forward
rapidity (with ca.~10(20)\% uncertainty). 
We utilize an effective $\Psi$-N absorption cross section, 
$\sigma_{\rm abs}$, to parameterize both nuclear shadowing and 
absorption. Applying Eq.~(\ref{glauber}) to $p$-$A$ collisions at SPS 
we obtain $\sigma_{\rm abs}^{J/\psi}$=7.3$\pm$1\,mb from the recent 
NA60 data at $E_{\rm lab}$=158\,GeV (corresponding to 
$\sqrt{s_{NN}}$=17.3\,GeV)~\cite{Arnaldi:2010ky}. The new experimental 
measurement at 158\,GeV turns out to give a significantly larger
value than previously available for 400\,GeV proton projectiles, 
$\sigma_{\rm abs}\simeq$4.4\,mb~\cite{Alessandro:2006jt} (the 
latter has been confirmed by NA60~\cite{Arnaldi:2010ky}), which
has been used in 
our previous calculations~\cite{Grandchamp:2003uw,Zhao:2007hh}. 
The comparison with recent PHENIX data~\cite{Adare:2007gn,Frawley-priv} 
yields $\sigma_{\rm abs}\simeq3.5$\,mb (5.5\,mb) for 
$\sqrt{s}$=200\,AGeV Au-Au collisions at mid rapidity, $|y|<0.35$ 
(forward rapidity, $|y|\in[1.2,2.2]$). For simplicity, we assume
the same absorption cross sections for the $\chi_c$ as for the $J/\psi$.
However, for excited states $\sigma_{\rm abs}$ is expected to be
significantly larger, even if they are not fully formed when the 
dissociation occurs. Taking guidance from the NA50 measurement with
400\,GeV protons, we use $\sigma_{\rm abs}^{\psi'}\simeq13$\,mb at 
$\sqrt{s}$=17.3\,AGeV and $\sigma_{\rm abs}^{\psi'}\simeq6.5(10)$\,mb 
at $\sqrt{s}$=200\,AGeV for mid (forward) $y$. 
For each charmonium state, the number surviving the pre-equilibrium 
stage in an $A$-$A$ collision at impact parameter $b$ thus amounts to 
\begin{eqnarray}
N_{\Psi}(b)&=&\int f_\Psi(\vec{b},\vec{x_t},\tau_0) d^2 \vec{x_t}.
\label{nucl_supp}
\end{eqnarray} 

The rather pronounced rapidity dependence of $\sigma_{\rm abs}$ at RHIC 
casts doubt on interpreting this quantity as an actual absorption cross 
section. It seems more reasonable to associate its increase at forward 
$y$ with nuclear shadowing~\cite{Ferreiro:2009ur} since the dissociation 
kinematics is very similar between mid and forward rapidity. While this 
does not affect the use of our ``effective" $\sigma_{\rm abs}$,
it does imply a nuclear shadowing effect on the open-charm cross
section in $A$-$A$ collisions (which is an important ingredient in the
calculation of regeneration). As a ``minimal" scheme we therefore 
associate the additional absorption of the $J/\psi$ yield at forward $y$ 
(relative to mid rapidity) with a suppression of open charm production
caused by shadowing, while we assume no shadowing corrections at mid
rapidity. Thus, at both SPS and RHIC the number of primordially 
produced $c\bar c$ pairs at mid-rapidity is calculated from 
the $pp$ cross section as
\begin{equation}
N^{\rm mid}_{c\bar c}(b)=
\left.\Delta y\frac{\dd\sigma^{c\bar c}_{pp}}{\dd y}\right\vert_{y=0}
T_{AB}(b) \ ,
\label{Ncc_mid}
\end{equation}
while for forward $y$ at RHIC we use 
\begin{equation}
N^{\rm for}_{c\bar c}(b)=
\left.\Delta y\frac{\dd\sigma^{c\bar c}_{pp}}{\dd y}\right\vert_{y=1.7} 
T_{AB}(b)\frac{S_{\rm nuc}^{\rm for}}{S_{\rm nuc}^{\rm mid}} \ . 
\label{Ncc_forw}
\end{equation}
Here, $T_{AB}(b)$ is the usual nuclear overlap function and 
$S_{\rm nuc}$ denotes the $J/\psi$ suppression factor due to CNM 
effects, parameterized by $\sigma_{\rm abs}$ in the Glauber formula, 
Eq.~(\ref{glauber}). In particular, the ratio 
$S_{\rm nuc}^{\rm for}/S_{\rm nuc}^{\rm mid}$ represents the extra 
suppression associated with nuclear shadowing, operative for both 
$J/\psi$ and $c\bar c$ production. The input charm-quark cross 
section in $pp$ is taken as $\dd \sigma_{\bar cc}/\dd y$
($y$=0)=2.2\,$\mu$b at SPS (according to the recent compilation 
of data in Ref.~\cite{Lourenco:2006vw}), and as 
$\dd \sigma_{\bar cc}/\dd y$ ($y$=0)=123$\pm$40\,$\mu$b at RHIC 
(in line with recent PHENIX measurements~\cite{Adare:2006hc}). At  
forward rapidity we assume the $pp$ charm-quark cross section to
be reduced by 1/3
according to recent measurements~\cite{Zhang:2008kr}.

\subsection{Fireball Model}
\label{ssec_fb}
\begin{figure}[!t]
\centering
\includegraphics[angle=-90,width=0.5\textwidth]{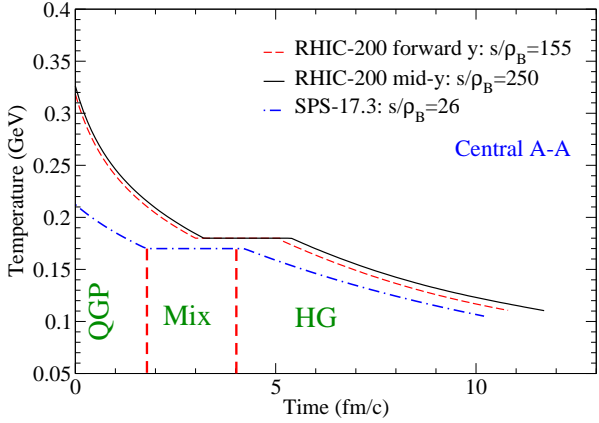}
\caption{(Color online) Time profiles of temperature for central 
  collisions of heavy nuclei (participant number $N_{\rm part}$=380) at 
  RHIC ($\sqrt{s}$=200\,AGeV; solid line: mid rapidity; dashed line: 
  forward rapidity) and SPS ($\sqrt{s}$=17.3\,AGeV; dot-dashed line).}
\label{fig_fb}
\end{figure}

Once the nuclear-collision system thermalizes, its temperature is the 
key quantity connecting to the in-medium properties of the charmonia 
as discussed in Sec.~\ref{sec_equil}. As in our previous 
work~\cite{Grandchamp:2001pf,Grandchamp:2003uw,Zhao:2007hh}, we 
estimate the time evolution of the temperature using an isentropically 
and cylindrically expanding isotropic fireball characterized by an 
eigenvolume,
\begin{equation}
\label{Vfb}
V_{\rm FB}(\tau)=(z_0+v_z\tau+\frac{1}{2}a_z \tau^2) \ \pi \ 
(R_0+\frac{1}{2}a_\perp \tau^2)^2\ .
\end{equation} 
The fireball expansion parameters ${v_z,a_z,a_{\perp}}$ are chosen
such that the hadron spectra at thermal freezeout are consistent with
the empirically extracted light-hadron flow in resemblance of
hydrodynamical calculations (we actually use a relativistic form of
$a_{\perp}(\tau)$ which limits the surface speed,
$v_s(\tau)$=$a_{\perp}\tau$, to below $c$). The initial transverse
radius $R_0$ represents the initial transverse overlap of the two
colliding nuclei at a given impact parameter $b$, while the initial
longitudinal length, $z_0$, is related to thermalization time $\tau_0$
through $z_0\simeq\Delta y\tau_0$ where $\Delta y$=1.8 represents the
typical longitudinal rapidity coverage of a thermal fireball. We
assume that at a formation time of $\tau_0=1.0(0.6)$\,fm/$c$ the
medium at SPS (RHIC) first thermalizes with all the entropy, $S_{\rm
  tot}(b)$, being built up. The latter is estimated from the
multiplicities of observed charged particles and assumed to be
conserved during the adiabatic expansion.  We can then use the entropy
density, $s(\tau)$=$S_{\rm tot}/V_{\rm FB}$ at each moment $\tau$ in
the evolution to infer the fireball temperature once we specify the
equation of state of the medium, specifically $s(T)$. The QGP is
modeled by an ideal gas of massive quarks and gluons while the
hadronic phase is approximated by a non-interacting resonance gas with
76 mesonic and baryonic states up to masses of 2\,GeV. The critical
temperature, $T_c$=170(180)\,MeV at SPS(RHIC) is roughly consistent
with thermal-model fits to observed particle
ratios~\cite{BraunMunzinger:2003zd} and predictions of lattice
QCD~\cite{Cheng:2006qk}. A freeze-out temperature of $T_{\rm
  fo}\simeq120$\,MeV terminates the evolution and results in a total
fireball lifetime of $\tau_{\rm fo}$=10-12\,fm/$c$ for central $A$-$A$
collisions. The resulting temperature evolution as a function of time
$\tau$ is displayed in Fig.~\ref{fig_fb} for SPS and RHIC. Note that
there is little difference between mid ($|y|<0.35$) and forward
rapidity ($|y|\in[1.2,2.2]$) for Au-Au collisions at RHIC due to the
slowly varying rapidity density of charged particles over this $y$
range~\cite{Arsene:2004fa}, cf.~Ref.~\cite{Zhao:2008pp} for more
details.

\subsection{Rate Equation in Hot Medium}
\label{ssec_rate}
We now proceed to the thermal rate equation to calculate the time 
dependence of the charmonium number, $N_{\Psi}$, throughout the third 
stage of the heavy-ion reaction,
\begin{equation}
\label{rate-eq}
\frac{\dd N_{\Psi}}{\dd \tau}=
-\Gamma^{\rm diss}_{\Psi}(T) \ [N_{\Psi}-N_{\Psi}^{\rm eq}(T) ] \ .
\end{equation}
The rate equation is solved separately for $\Psi$=$J/\psi$, $\chi_c$
and $\psi'$, with initial conditions given by Eq.~(\ref{nucl_supp}). 
The loss term, $-\Gamma^{\rm diss}_{\Psi}N_{\Psi}$, accounts for the
dissociation of primordially produced charmonia and the gain term, 
$\Gamma^{\rm diss}_{\Psi}N_{\Psi}^{\text{eq}}$, for the regeneration
of charmonia via coalescence of $c$ and $\bar c$ quarks. Both processes
are governed by the temperature-dependent inelastic reaction rate, 
$\Gamma^{\rm diss}_{\Psi}(T)$, which is taken from Sec.~\ref{ssec_diss}
(note that we do not employ the $p$=0 value of the rate but rather
a 3-momentum averaged value which is slightly larger; its precise value
is obtained by matching the final yield of the loss term to the exact
result obtained from solving the momentum-dependent Boltzmann equation 
(\ref{boltz}) below). 
The equilibrium limit of each charmonium state, $N_{\Psi}^{\rm eq}(T)$,
is intricately related to their equilibrium properties, which will be
discussed further below. 
To make the decomposition of the $J/\psi$ number at any time $\tau$,
\begin{equation}
N_{\Psi}(\tau)=N^{\rm prim}_{\Psi}(\tau)+N^{\rm reg}_{\Psi}(\tau) \ , 
\label{Npsi-tau}
\end{equation}
into (suppressed) primordial charmonia, $N^{\rm prim}_{\Psi}$,  and
regenerated ones, $N^{\rm reg}_{\Psi}(\tau)$, more explicit, we 
exploit the linearity of the rate equation~(\ref{rate-eq}). 
We define $N^{\rm prim}_{\Psi}(\tau)$ as the solution of the 
homogeneous rate equation,
\begin{equation}
\label{rate-eq_dir}
\frac{\dd N_{\Psi}^{\rm prim}}{\dd \tau}=
-\Gamma^{\rm diss}_{\Psi} \ N_{\Psi}^{\rm prim} \ ,
\end{equation}
with the same initial condition as for the full rate equation,
$N_{\Psi}^{\rm prim}(0)=N_{\Psi}(0)$.  The regeneration component, 
$N^{\rm reg}_{\Psi}$, then follows as the difference between the 
solution of the full and the homogeneous rate equation,
which can be expressed as 
\begin{equation}
\label{rate-eq_reg}
\frac{\dd N_{\Psi}^{\rm reg}}{\dd \tau}=
-\Gamma^{\rm diss}_{\Psi} \ (N_{\Psi}^{\rm reg}-N_{\Psi}^{\rm eq}) \ 
\end{equation} 
with vanishing initial condition, 
$N_{\Psi}^{\rm reg}(\tau<\tau_0^\Psi)=0$. The onset time of 
regeneration processes, $\tau_0^\Psi$, is 
defined by $T(\tau_0^\Psi)= T^{\rm diss}_\Psi$ for each state $\Psi$.


Let us now return to the equilibrium limit of the charmonium abundances,
$N_{\Psi}^{\rm eq}(T)$, which we evaluate within the statistical model.
Since the thermal production and annihilation rates of $c\bar c$ are 
believed to be small at SPS and RHIC energies, $c\bar c$ pairs are 
assumed to be exclusively produced in primordial $N$-$N$ collisions
and conserved thereafter. The open and hidden charm states are then
populated in relative chemical equilibrium according to the
canonical charm-conservation equation,
\begin{equation}
N_{c\bar c}=\frac{1}{2}~N_{\rm op}~\frac{I_1(N_{\rm op})}
{I_0(N_{\rm op})} + N_{\rm hid} \ ,
\label{Ncc}
\end{equation}
with $N_{c\bar c}$: total number of charm-quark pairs from initial 
production, $N_{\rm op} =\gamma_c V_{\rm FB} n_{\rm op}$: total number 
of all open-charm states with pertinent equilibrium density $n_{\rm op}$,
$N_{\rm hid}=\gamma_c^2 V_{\rm FB} n_{\rm hid}$: total number of all 
charmonium states with pertinent equilibrium density $n_{\rm hid}$,
and $\gamma_c$: charm-quark fugacity accounting for the deviation of 
chemical equilibrium with the heat bath ($\gamma_c$=1 in full 
equilibrium). The ratio of modified Bessel functions, 
$I_1(N_{\rm op})/I_0(N_{\rm op})$, on the right-hand-side of 
Eq.~(\ref{Ncc}) is the characteristic canonical suppression 
factor which accounts for the exact conservation of net-charm
number, $N_c-N_{\bar c}$, in the limit of small $N_{c\bar c}$
(canonical limit)~\cite{Cleymans:1990mn,Gorenstein:2000ck}: for 
$N_{\rm op}\ll 1$, one has 
$I_1(N_{\rm op})/I_0(N_{\rm op})\to\frac{1}{2}N_{\rm op}$, 
which acts as an additional (small) probability to enforce a
vanishing net charm content in the system (\ie, both $c$ and
$\bar c$ have to be present simultaneously). 

The open-charm number, $N_{\rm op}$, is evaluated as follows.  For the
QGP phase in the weak-binding scenario only charm quarks are counted
as open-charm states. In the strong-binding scenario, the $T$-matrix
calculations of Ref.~\cite{Riek:2010fk} suggest that $c\bar q$ and
$\bar cq$ (charm-light) bound states survive in QGP up to
$\sim$1.3\,$T_c$; therefore, we count both charm quarks and the
lowest-lying $S$-wave $D$-mesons ($D$, $D^*$, $D_s$ and $D^*_s$), as
open charm states for $T<1.3T_c$. The charm-quark masses in the QGP
correspond to the temperature-dependent ones displayed in
Fig.~\ref{fig_mc}, while for the meson resonances above $T_c$ we
estimate from Ref.~\cite{Riek:2010fk} $m_{D}=m_{D^*}\simeq2.0$\,GeV
and $m_{D_s}=m_{D^*_s}\simeq2.1$\,GeV (hyperfine splitting has been
neglected).  For the HG phase all charmed hadrons listed by the
particle data group~\cite{Yao:2006px} are counted as open-charm
states, with their vacuum masses.  The hidden charm number, $N_{\rm
  hid}$, is evaluated in line with the existing charmonium states and
their masses at given temperature $T$, but its contribution to
$N_{c\bar c}$ is numerically negligible.

\begin{figure}[!t] 
\centering
\includegraphics[angle=-90,width=0.48\textwidth]{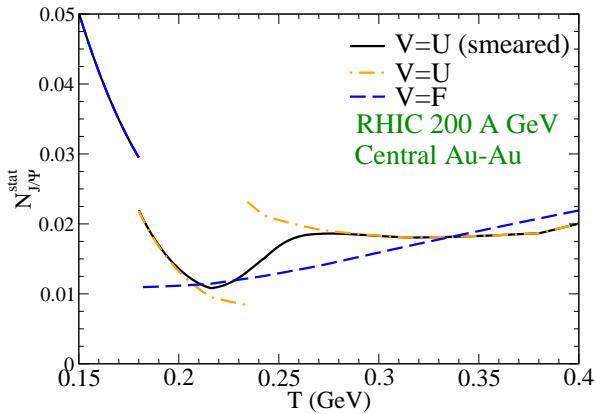}
\caption{(Color online) Temperature dependence of the in-medium 
   $J/\psi$ equilibrium limit using the statistical model in the
   QGP within the strong-binding scenario (dot-dashed lines: 
   with and without $D$-meson resonances below and above
   1.3\,$T_c$$\simeq$234\,MeV, respectively; solid line: smooth
   interpolation of the previous two cases; see text for
   details), the weak-binding scenario (dashed line) and
   in the HG for temperatures below $T_c$=180\,MeV.}
\label{fig_jpsi-eq}
\end{figure}

Knowing $n_{\rm op}$, $n_{\rm hid}$ and $V_{\rm FB}$ at each 
temperature, one can solve Eq.~(\ref{Ncc}) for the charm-quark 
fugacity, $\gamma_c(T)$, and apply it to compute the statistical
equilibrium limit of each charmonium state as
\begin{equation} 
N_{\Psi}^{\rm stat}=\gamma_c^2 \ V_{\rm FB} \ n_{\Psi}
\label{Npsi-stat}
\end{equation} 
in terms of its equilibrium density, $n_{\Psi}$.
In Fig.~\ref{fig_jpsi-eq} we collect the numerical results of the 
statistical equilibrium limit for $J/\psi$ abundances (excluding 
feeddown) for central 200\,AGeV Au-Au collisions at RHIC.
The discontinuity at 1.3\,$T_c$ for the strong-binding scenario
(dot-dashed line) is due to the inclusion of the $D$ resonances in 
the QGP medium. We smoothly interpolate around the melting temperature
for the $D$-mesons with a hyperbolic tangent function (solid line) 
to represent a more gradual (dis)appearance of the $D$ resonances
(we have checked that this procedure has negligible impact on 
the calculation of observables in Sec.~\ref{sec_data}). 

To achieve a more realistic implementation of the statistical 
equilibrium limit, we apply two corrections to $N_{\Psi}^{\rm stat}$ 
to schematically implement off-equilibrium effects of charm quarks
in momentum and coordinate space. The former is aimed at simulating 
incomplete thermalization of the charm-quark $p_t$ spectra throughout 
the course of the thermally evolving bulk medium. It is expected that 
the coalescence rate from non- or partially thermalized $c$- and 
$\bar c$-quark spectra is smaller than for fully thermalized 
ones~\cite{Grandchamp:2002wp,Greco:2003vf}, since the former are 
harder than the latter
and thus provide less phase-space overlap for charmonium bound-state
formation. We implement this correction by multiplying the charmonium 
abundances from the statistical model with a schematic relaxation
factor~\cite{Grandchamp:2002wp}, 
\begin{equation}
N_{\Psi}^{\rm eq}={\mathcal R}(\tau) \  N_{\Psi}^{\rm stat} \ , \ 
{\mathcal R}(\tau)=1-\exp (-\tau/\tau^{\rm eq}_c)
\end{equation}
where $\tau^{\rm eq}_c$ is a parameter which qualitatively represents
the thermal relaxation time of charm quarks (it is one of our 2 main
adjustable parameters in our phenomenological applications in
Sec.~\ref{sec_data}). A rough estimate of this time scale may be
obtained from microscopic calculations of this quantity within the
same $T$-matrix approach as used here for charmonia, where the thermal
charm-quark relaxation time turns out to be $\tau_{\rm
  eq}^c\simeq$3-10\,fm/$c$~\cite{vanHees:2007me,Riek:2010fk}.  Such
values allows for a fair description of open heavy-flavor suppression
and elliptic flow at RHIC~\cite{vanHees:2005wb,vanHees:2007me}.  The
second correction is applied in coordinate space, based on the
realization that, after their pointlike production in hard $N$-$N$
collisions, the $c$ and $\bar c$ quarks only have a limited time to
diffuse throughout the fireball volume. At RHIC and especially at SPS
only few $c\bar c$ pairs are produced (\eg, $\dd N_{c\bar
  c}/dy\simeq1.2$ in semicentral ($b$=7\,fm) Au-Au collisions at
RHIC), and the hadronization time is smaller than the fireball
radius. Thus, $c$ and $\bar c$ will not be able to explore the full
fireball volume but rather be restricted to a ``correlation volume'',
$V_{\rm corr}$~\cite{Heinz:2001priv,Grandchamp:2003uw} (the analogous
concept has been successfully applied to strangeness production in
$p$-$A$ and $A$-$A$ collisions in the SPS energy
regime~\cite{Hamieh:2000tk}).  We implement this correction by
replacing the fireball volume $V_{\rm FB}$ in the argument of the
Bessel functions in Eq.~(\ref{Ncc}) by the correlation volume $V_{\rm
  corr}$~\cite{Grandchamp:2003uw,Young:2008he}. The latter is
identified with the volume spanned by a receding $c\bar c$ pair,
\begin{equation}
V_{\rm corr}(\tau)=\frac{4\pi}{3} (r_0 + \langle v_c\rangle \tau)^3 \ ,
\end{equation}
where $r_0\simeq1.2$\,fm represents an initial radius characterizing
the range of strong interactions, and $\langle v_c\rangle$ is an
average speed with which the produced $c$ and $\bar c$ quark recede
from the production point; we estimate it from the average $p_t$ in
$D$-meson spectra in $p$-$A$
collisions~\cite{Klinksiek:2005kq,Grandchamp:2003uw,Rapp:2009my} as
$\langle v_c\rangle\simeq0.55(0.6)c$ at SPS (RHIC).  The correlation
volume leads to a significant increase of $\gamma_c$ (since $I_0/I_1$
is reduced) and thus of the modified $\Psi$ ``equilibrium limit" due
to an effectively larger $c\bar c$ density.

\subsection{Transverse-Momentum Dependence}
\label{ssec_pt-dep}
The transverse momentum dependence of in-medium charmonium production in 
$A$-$A$ collisions has been suggested as a tool to better discriminate 
primordial and regenerated production~\cite{Grandchamp:2001pf}. 
To investigate this observable within our approach we adopt the 
procedure suggested in Ref.~\cite{Zhao:2007hh} where the $p_t$
spectra for the 2 components are evaluated at hadronization based on 
the decomposition given by Eqs.~(\ref{Npsi-tau})-(\ref{rate-eq_reg}).
For the primordial component we employ a more differential version of 
the rate equation~(\ref{rate-eq_dir}), \ie, a Boltzmann transport 
equation, to describe the evolution of the charmonium phase-space 
distribution functions, $f_\Psi(\vec{x},\vec{p},\tau)$, in a
thermalized medium,
\begin{equation}
p^{\mu} \, \partial_{\mu}f_\Psi^{\rm prim}(\vec{x},\vec{p},\tau)
=-E_\Psi \ \Gamma^{\rm diss}_\Psi(\vec{x},\vec{p},\tau) \ 
f_\Psi^{\rm prim}(\vec{x},\vec{p},\tau) \ 
\label{boltz}
\end{equation}
where $p_0=E_\Psi=(m_\Psi^2+\vec p^{\,2})^{1/2}$.
The initial condition, $f^{\rm prim}_\Psi(\vec{x},\vec{p},\tau_0)
=f_\Psi(\vec{x},\tau_0) \ f_\Psi(\vec{p},\tau_0)$, is still given
as discussed in Sec.~\ref{ssec_cnm}. As mentioned before, we neglect 
elastic charmonium rescattering, and consequently
the modification of the charmonium $p_t$-spectra in the thermal
stage is mostly due to the momentum dependence of the dissociation 
rates, $\Gamma^{\rm diss}_{\Psi}(\vec{x},\vec{p},\tau)$, given in 
Eq.~(\ref{rate}) and plotted in the lower panel of 
Fig.~\ref{fig_diss-rate}. 
We also account for ``leakage" effects, \ie, charmonia escaping
the fireball volume are no longer subject to suppression, which
leads to the opposite trend compared to the increased suppression
with $p$ induced by the momentum dependence of the dissociation rate
(charmonia with low $p_t$ are more strongly suppressed since they 
stay longer within the fireball). 
At the moment of freeze-out, $\tau_{\rm fo}$, we obtain the final 
$p_t$ spectra of the primordial component by integrating the 
spatial part of the solution of the Boltzmann equation,
\begin{equation}
\frac{\dd N_{\Psi}^{\rm prim}}{p_t\dd p_t}
=2\pi\int d^2x f_\Psi(\vec{x},\vec{p_t},\tau_{\rm fo}) \ .
\label{pt-dir}
\end{equation}

A microscopic evaluation of the momentum dependence of the
regeneration component using the quasifree mechanism requires to 
compute a 3-to-2 process, $i+c+\bar c\to i + \Psi $. In the framework 
of the Boltzmann equation, this is a rather involved calculations 
which will be reported in an upcoming publication~\cite{Zhao:2010-2}.
For the time being, we approximate the $p_t$ distributions of 
regenerated charmonia by local thermal distributions boosted 
by the transverse flow of the medium, amounting to a standard 
blastwave description~\cite{Schnedermann:1993ws},
\begin{equation}
\label{pt-reg}
\frac{\dd N_{\Psi}^{\rm reg}}{p_t\dd p_t}\propto 
m_t\int^R_0 rdr K_1\left(\frac{m_t\cosh y_t}{T}\right)
I_0\left(\frac{p_t\sinh y_t}{T}\right) \ 
\end{equation}
($m_t=\sqrt{m_\Psi^2+p^2_t}$). The medium is characterized by the 
transverse-flow rapidity $y_t=\tanh^{-1}v_t(r)$ using a linear flow 
profile $v_t(r)=v_{s}\frac{r}{R}$ with a surface velocity 
$v_s=a_\perp\tau_{\rm mix}$ and transverse fireball radius $R=R(\tau_{\rm mix})$ as given by the fireball expansion formula, 
Eq.~(\ref{Vfb}), at the end of the mixed phase, $\tau_{\rm mix}$. 
We evaluate the blastwave expression at the hadronization transition 
($T_c$) and neglect rescattering of $\Psi$'s in the hadronic phase.  
Two additional effects are neglected in this treatment which
to a certain extent tend to compensate each other: 
on the one hand, due to incomplete charm-quark thermalization, one 
expects the regenerated charmonium spectra to be harder than in 
equilibrium, but, on the other hand, a good part of the regeneration 
occurs before the mixed phase~\cite{Rapp:2005rr,Yan:2006ve} (see also
Fig.(\ref{jpsi-time-evo}) below) so that the evaluation of the blastwave 
expression at the end of the mixed phase presumably overestimates the 
blue shift due to the flow field. An explicit evaluation of the gain 
term with realistic (time-dependent) charm-quark spectra within a 
Boltzmann transport equation~\cite{Zhao:2010-2} will be able to
lift these approximations. 

The total charmonium $p_t$ spectra are the sum of the suppressed 
primordial and regenerated parts, Eqs.~(\ref{pt-dir}) and 
(\ref{pt-reg}), with an absolute normalization following from the
decomposition of the $p_t$-integrated rate equation, 
Eq.~(\ref{Npsi-tau}), at thermal freezeout.
The average $p_t$ squared is readily computed as
\begin{equation} 
\langle p_t^2 \rangle^{\rm prim,reg}=
\frac{\int d^2p_t \ p^2_t \ \frac{\dd N_{\Psi}^{\rm prim,reg}}
{p_t\dd p_t}}
{\int d^2p_t \ \frac{\dd N_{\Psi}^{\rm prim,reg}}{p_t\dd p_t}} \ ,
\label{pt2}
\end{equation}
which we compare to experimental data as part of the following section.

\section{Comparison to SPS and RHIC Data}
\label{sec_data}
In this section we present and discuss the numerical applications
of the above framework to $J/\psi$ data in URHICs at SPS and RHIC.
For each observable, we confront the results of the strong- and 
weak-binding scenario in an attempt to discriminate qualitative
features. The feeddown to $J/\psi$ from $\chi_c$ and $\psi'$ states 
is taken into account, assuming fractions of 32\% and 8\%, 
respectively, for primordial production in $pp$ collisions.  
We have divided the discussion into the centrality dependence
of inclusive $J/\psi$ yields in Sec.~\ref{ssec_incl} and
their $p_t$ dependence in Sec.~\ref{ssec_pt-spec}.

\subsection{Centrality Dependence of Inclusive Yields}
\label{ssec_incl}
The $J/\psi$ yield in $A$-$A$ collisions is usually quantified
in terms of the nuclear modification factor as a function
of centrality,
\begin{equation}
\label{raa}
R_{AA}(b)=\frac {N_{J/\psi}^{AA}(b)}{N_{J/\psi}^{pp}N_{\rm coll}(b)} \ ,
\end{equation}
where $N_{\rm coll}(b)$ is the number of binary collisions of the
incoming nucleons at impact parameter $b$.
Before we turn to the results, we recall the two main parameters in 
our approach, which are the strong coupling constant, $\alpha_s$,
the thermal charm-quark relaxation time, $\tau_c^{\rm eq}$
The former controls the inelastic charmonium reaction rate and the 
latter the magnitude of the $\Psi$ equilibrium limits. We adjust
them to approximately reproduce the inclusive $J/\psi$ yield for 
central $A$-$A$ collisions at SPS and RHIC, within reasonable bounds. 
For $\alpha_s$ we find that a common value of 0.32, which is at the 
upper end of the value in the Coulomb term in the $Q\bar Q$ free 
energy, can be used, in combination with $\tau_c^{\rm eq}$=3.8\,fm/$c$ 
for the strong-binding scenario and $\tau_c^{\rm eq}$=1.6\,fm/$c$ for 
the weak-binding scenario. For simplicity, we refrain from introducing 
an additional temperature dependence into these parameters. The
composition of the total yield, its centrality dependence and $p_t$ 
spectra can then be considered as a prediction within each of
the 2 scenarios. 

\begin{figure}[!t]
\centering
\includegraphics[angle=-90,width=0.48\textwidth]
{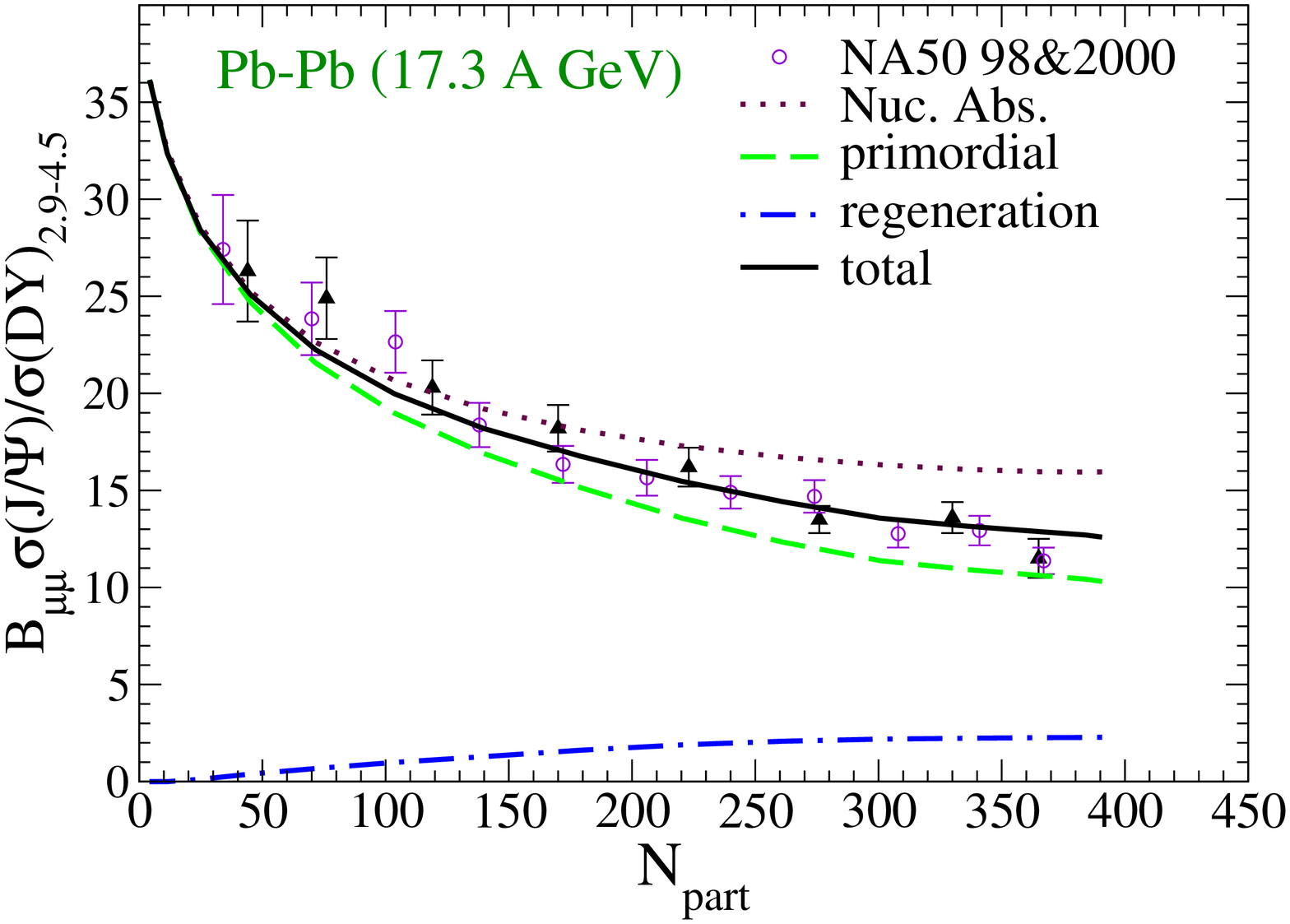}

\vspace{-0.8cm}

\includegraphics[angle=-90,width=0.48\textwidth]
{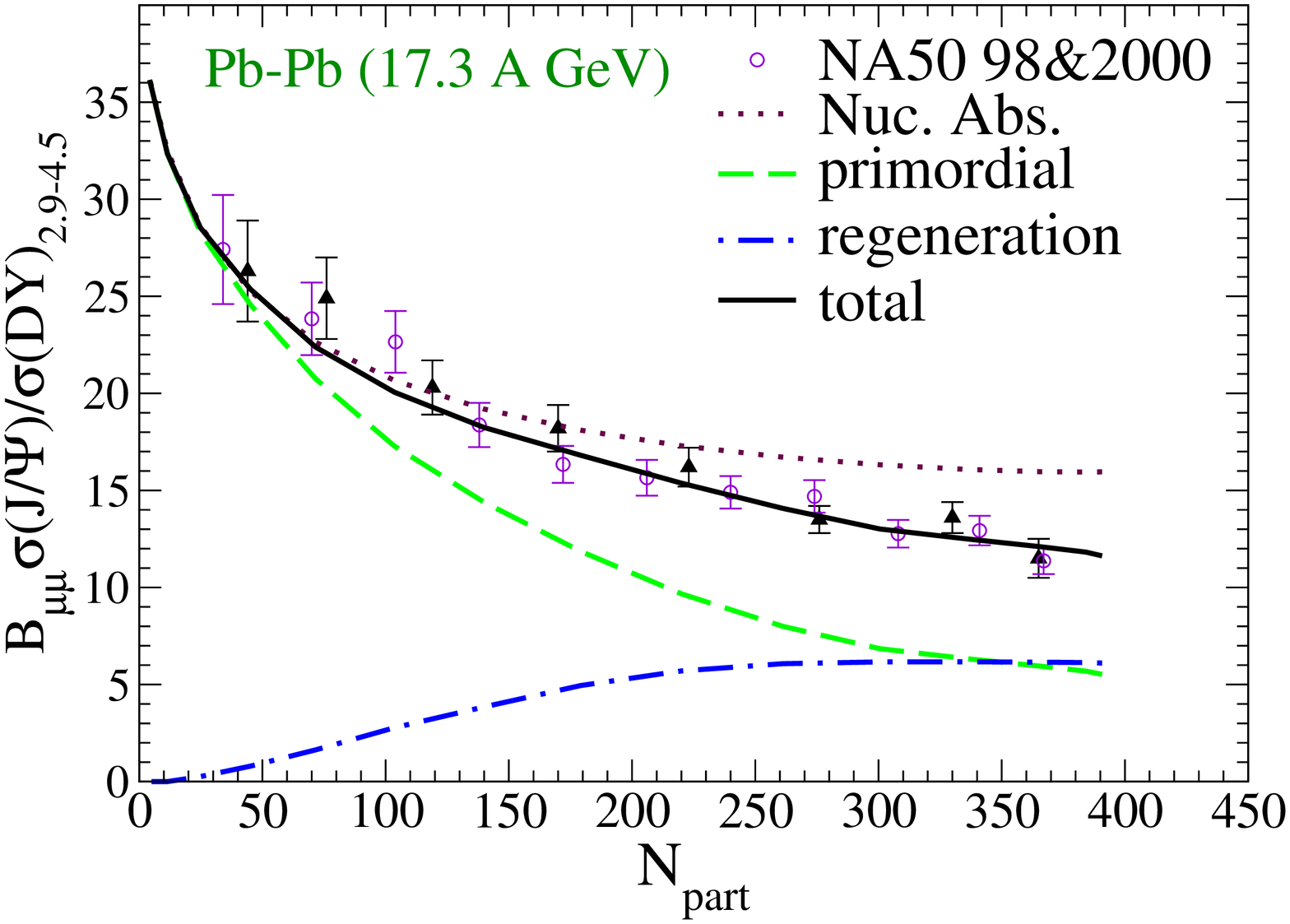}
\caption{(Color online) Results of the thermal rate-equation approach
  for $J/\psi$ production (normalized to Drell-Yan pairs) versus 
  centrality at SPS, compared to NA50 
  data~\cite{Ramello:2003ig,Alessandro:2004ap}. Solid lines: total
  $J/\psi$ yield; dashed lines: suppressed primordial production;
  dot-dashed lines: regeneration component; dotted lines: primordial
  production with CNM effects only. Upper panel: strong-binding
  scenario; lower panel: weak-binding scenario.}
\label{fig_raa-sps}
\end{figure}
\begin{figure}[!t]
\centering
\includegraphics[angle=-90,width=0.48\textwidth]
{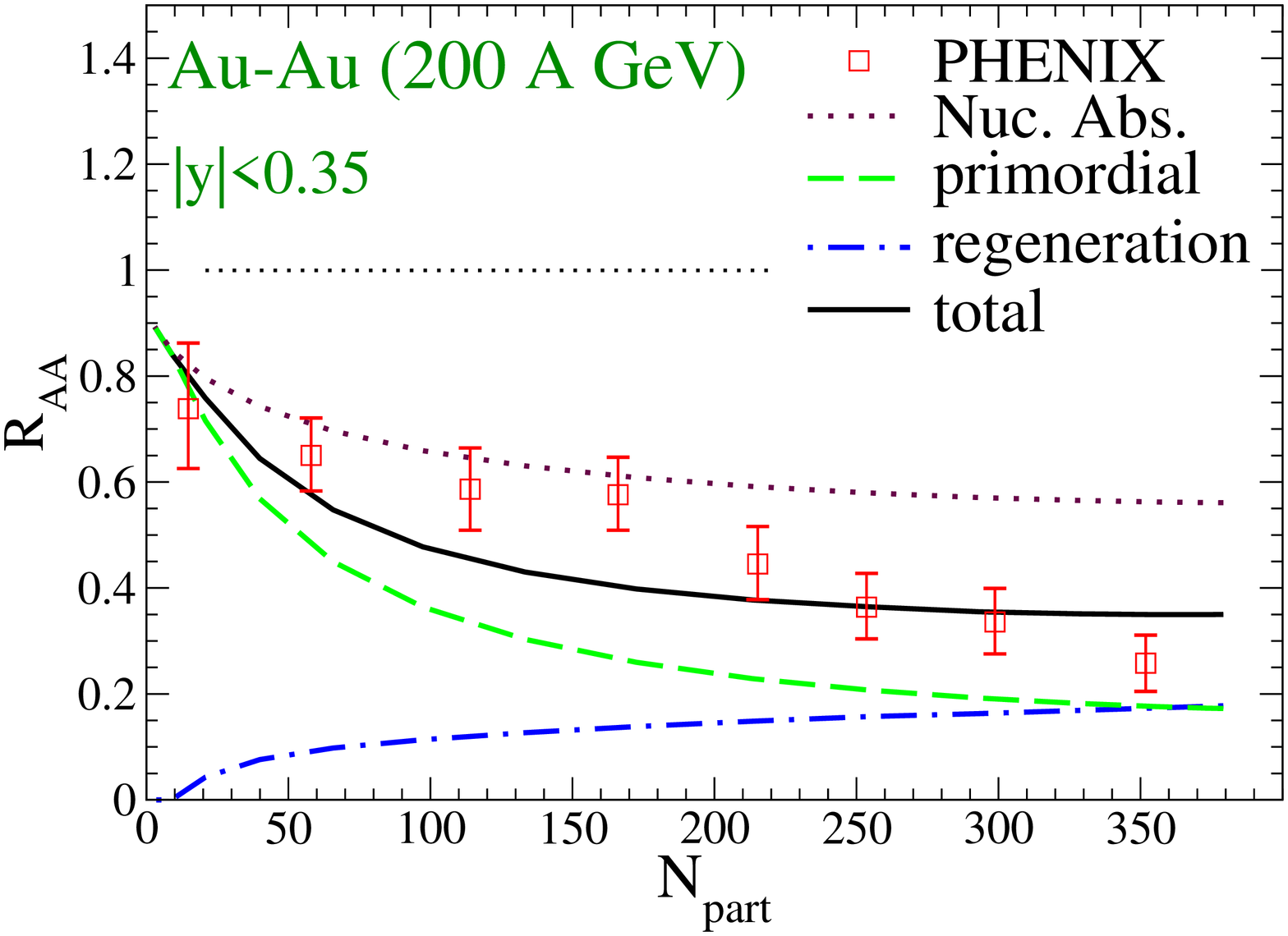}
                                                                          
\vspace{-0.8cm}
                                                                          
\includegraphics[angle=-90,width=0.48\textwidth]
{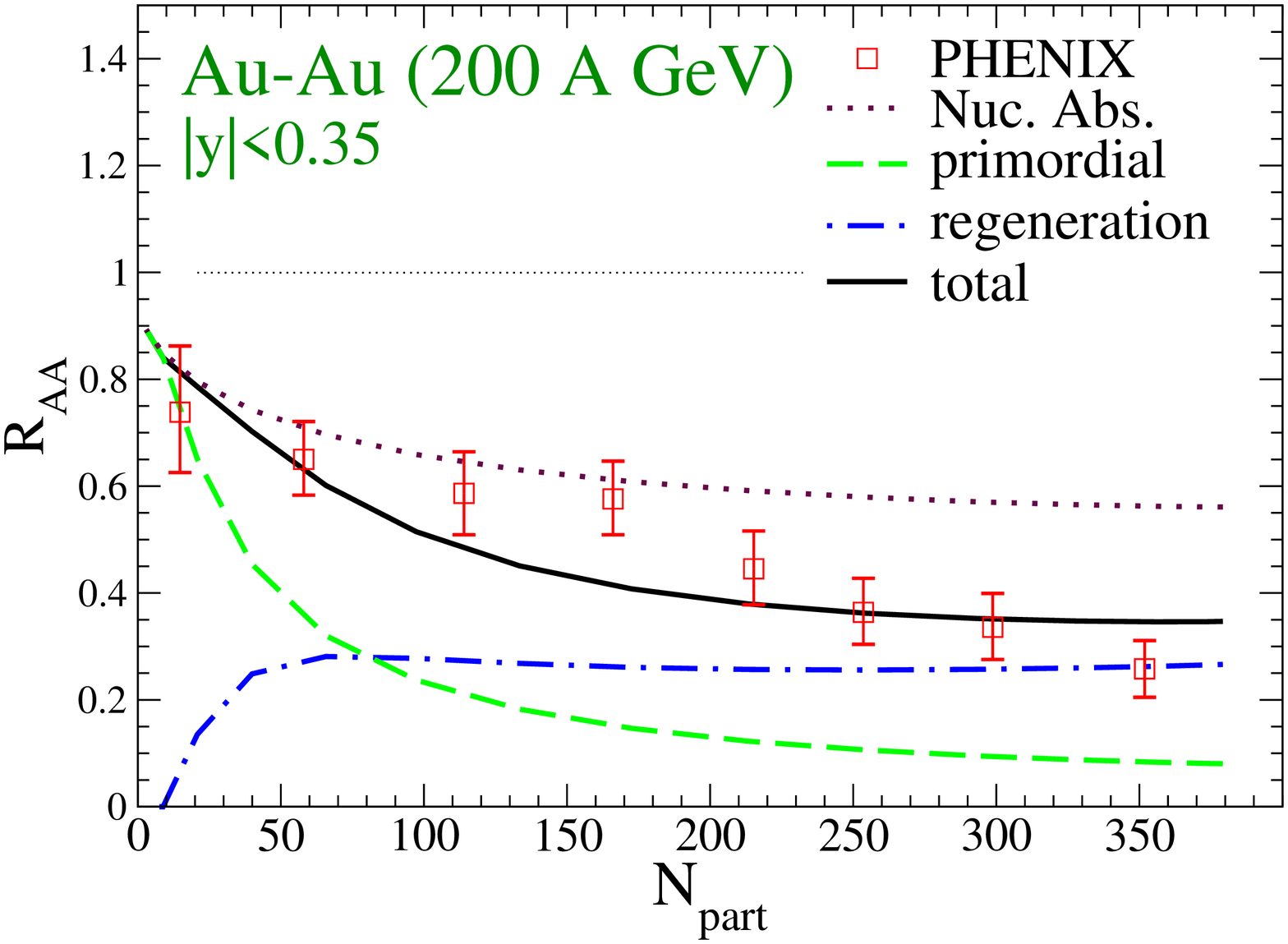}
\caption{(Color online) Results of the thermal rate-equation approach
  for the nuclear modification factor versus centrality at mid rapidity
  at RHIC, compared to PHENIX data~\cite{Adare:2006ns}.
  Solid lines: total $J/\psi$ yield; dashed lines: suppressed primordial
  production; dot-dashed lines: regeneration component; dotted lines:
  primordial production with CNM effects only. Upper panel:
  strong-binding scenario; lower panel: weak-binding scenario.}
\label{raa_rhic_mid}
\end{figure}

We begin with $J/\psi$ production in $\sqrt s$=17.3\,AGeV Pb-Pb
collisions at SPS, for which we compare our results in the strong-
and weak-binding scenario with NA50 data in Fig.~\ref{fig_raa-sps}.
For these data, the denominator in Eq.~(\ref{raa}) is replaced by
the number of Drell-Yan dileptons at high mass, while
the numerator includes the branching ratio into dimuons.
The pertinent proportionality factor, equivalent to the $pp$ limit
of this ratio (47.0$\pm$1.4~\cite{Roberta-priv}), and the CNM-induced
suppression (dotted line in Fig.~\ref{fig_raa-sps}) are
inferred from latest NA60 $p$-$A$ measurements~\cite{Arnaldi:2010ky},
which we reproduce using the Glauber model formula, Eq.~(\ref{glauber}),
with $\sigma_{\rm abs}$=7.3\,mb. The suppression of the primordial
component (dashed line) relative to the nuclear absorption
(dotted line) represents the ``anomalous'' suppression by the hot
medium, which increases with centrality due to higher initial 
temperatures and longer fireball lifetimes. The regeneration 
component increases with centrality as well, mostly due to
the increase of the $\cal{R}$-factor and the larger lifetime
which facilitates the approach to the equilibrium limit 
according to Eq.~(\ref{rate-eq_reg}). 
According to detailed balance between dissociation and regeneration, 
an increase in the former also implies an increase in the latter. 
The sum (solid line) of primordial and regeneration contributions
describes the centrality dependence of the inclusive $J/\psi$ yield 
at SPS reasonably well in both scenarios. In the strong-binding 
scenario the primordial component is dominant and the majority of 
the anomalous suppression originates from the dissociation of 
$\chi_c$ and $\psi'$, since at the temperatures realized at SPS 
($T\simeq200$\,MeV) the quasifree dissociation rates for $\chi_c$ 
and $\psi'$ are much larger than those for $J/\psi$, recall 
Fig.~\ref{fig_diss-rate}. In the weak-binding scenario, however, the 
regeneration yield becomes comparable to the primordial one for 
semi-/central collisions due to larger dissociation rates and the 
smaller charm-quark equilibration time scale.

\begin{figure}[!t]
\includegraphics[angle=-90,width=0.48\textwidth]
{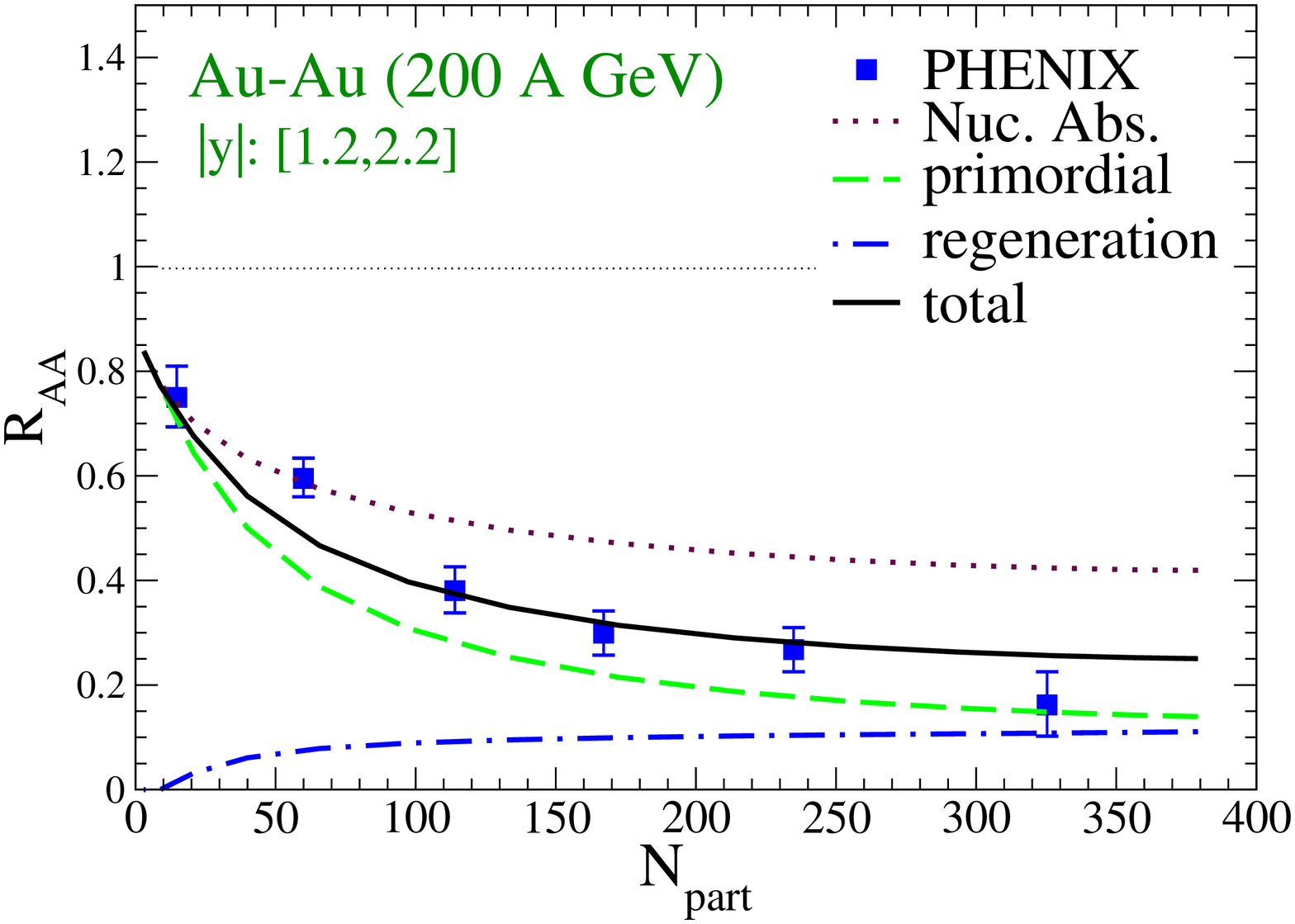}
                                                                         
\vspace{-0.8cm}
                                                                         
\includegraphics[angle=-90,width=0.48\textwidth]
{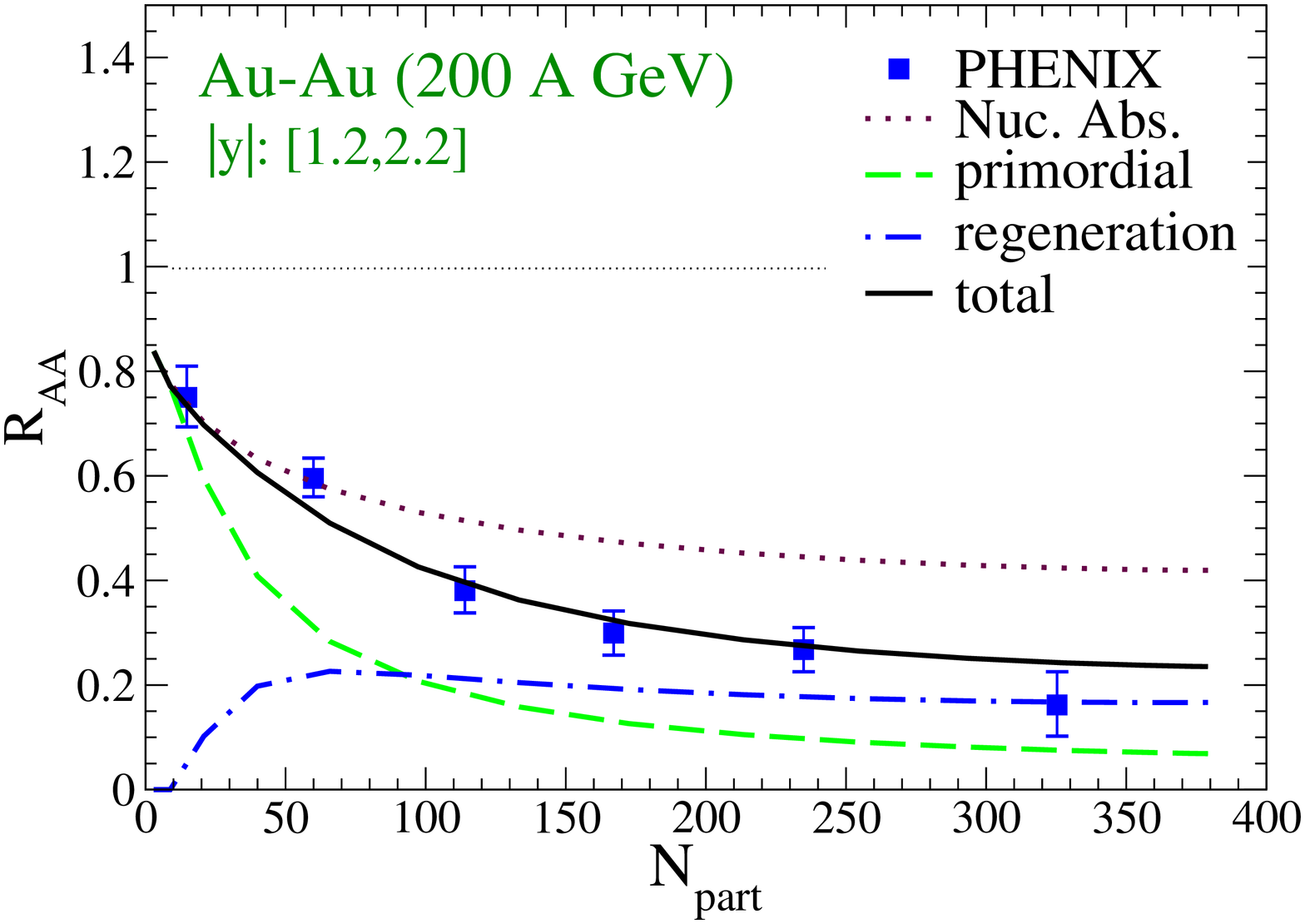}
\caption{(Color online) Results of the thermal rate-equation approach 
  for the nuclear modification factor  vs. centrality
  at forward rapidity compared to PHENIX
  data~\cite{Adare:2006ns}. Solid lines: total $J/\psi$ yield; dashed
  lines: suppressed primordial production; dot-dashed lines:
  regeneration component; dotted lines: primordial production with CNM
  effects only. Upper panel: strong-binding scenario; lower panel:
  weak-binding scenario.}
\label{raa_rhic_forw}
\end{figure}
\begin{figure}[!t]
\centering
\includegraphics[angle=-90,width=0.48\textwidth]
{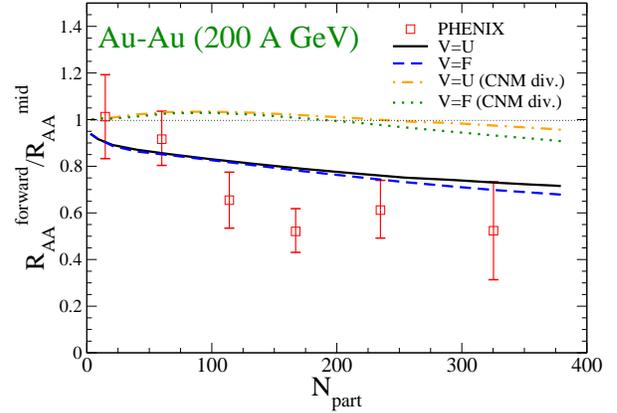}
\caption{(Color online) Results of the thermal rate-equation approach
  for the ratio of $R_{AA}$ for $J/\psi$ at forward and mid rapidity
  versus $N_{part}$ in strong (solid line) and weak (dashed line)
  binding scenarios compared to PHENIX data~\cite{Adare:2006ns}. In
  the upper two curves, CNM effects have been divided out in both
  numerator and denominator of the ratio.}
\label{raa_centra_forw_mid}
\end{figure}

Next we examine the centrality dependence of $J/\psi$ production
in 200\,AGeV Au-Au collisions at RHIC, first focusing on mid
rapidity ($|y|<0.35$), as shown in Fig.~\ref{raa_rhic_mid}.  The
suppression due to CNM effects (dotted line in Fig.~\ref{raa_rhic_mid})
is inferred from latest PHENIX d-Au measurements, which we reproduce
using the Glauber model formula Eq.~(\ref{glauber}) with
$\sigma_{\rm abs}$=3.5\,mb. For $N_{\rm part}\simeq0-100$, the
composition of primordial and regeneration contributions is
quite comparable to the SPS for $N_{\rm part}\simeq0-400$ within
both scenarios. Beyond $N_{\rm part}\simeq100$, suppression and
regeneration continue to increase, leveling off at an approximately
50-50\%  (20-80\%) partition for primordial and regeneration in the
strong-binding (weak-binding) scenario in central collisions.

Let us now turn to $J/\psi$ production at forward rapidity
($|y|\in[1.2,2.2]$) at RHIC, shown in Fig.~\ref{raa_rhic_forw}.
Again, both strong- and weak-binding scenarios reproduce the
experimental data fairly well, with similar relative partitions 
for primordial and regeneration contributions as at mid rapidity.
However, one of the ``puzzles" about $J/\psi$ production at RHIC
is the fact that the total $J/\psi$ yield is more strongly 
suppressed at forward rapidity than at mid rapidity. In our approach 
this follows from the stronger shadowing at forward rapidity leading to
less primordial production for both $J/\psi$ and $c\bar c$ pairs. The
former (latter) leads to a reduction of the direct (regeneration)
component. Since the thermodynamic properties of the fireball are
quite similar at mid and forward rapidity (recall Fig.~\ref{fig_fb}), 
charmonium suppression and regeneration in the hot medium are 
very similar between the mid and forward rapidity as discussed in 
Ref.~\cite{Zhao:2008pp}.  
To quantify the difference at forward rapidity and mid rapidity 
we display the ratio between the corresponding $R_{AA}$'s in 
Fig.~\ref{raa_centra_forw_mid} for both scenarios, which clearly
illustrates the importance of CNM effects to properly reproduce 
the data.

In Sec.~\ref{ssec_corr} we have argued that the strong- and weak-binding
scenarios discussed here may be considered as limiting cases 
for $J/\psi$ binding in the QGP, as bracketed by the identification
of the heavy-quark internal and free energies with a $Q$-$\bar Q$
potential. From the results above we believe that these scenarios
also provide a reasonably model-independent bracket on the role
of suppression and regeneration effects, in the following sense:   
At SPS, the strong-binding scenario defines a ``minimal" amount
of dissociation required to provide the anomalous suppression beyond
CNM effects (a small regeneration component is inevitable due to
detailed balance). The application to RHIC energy then implies 
an approximately equal partition of primordial and regenerated
charmonia in central Au-Au, not unlike Ref.~\cite{Yan:2006ve} where
the vacuum charmonium binding energies (``strong binding") have
been used in the QGP (together with the gluo-dissociation). 
On the other hand, in the weak-binding scenario, a large part of the 
$J/\psi$ yield in central $A$-$A$ is due to regeneration even at SPS, 
limited by the constraint that for sufficiently peripheral collisions 
(and at sufficiently large $p_t$) a transition to primordial production 
compatible with $p$-$A$ data should be restored. Clearly, for 
central $A$-$A$ at RHIC (and certainly at LHC) the final yield is 
then dominated by regeneration. 
Since both scenarios describe the inclusive yields reasonably well,
it is mandatory to investigate more differential observables to find
discriminating evidence which will be pursued in the following
section.  

\begin{figure}[!t]
\centering
\includegraphics[angle=-90,width=0.48\textwidth]
{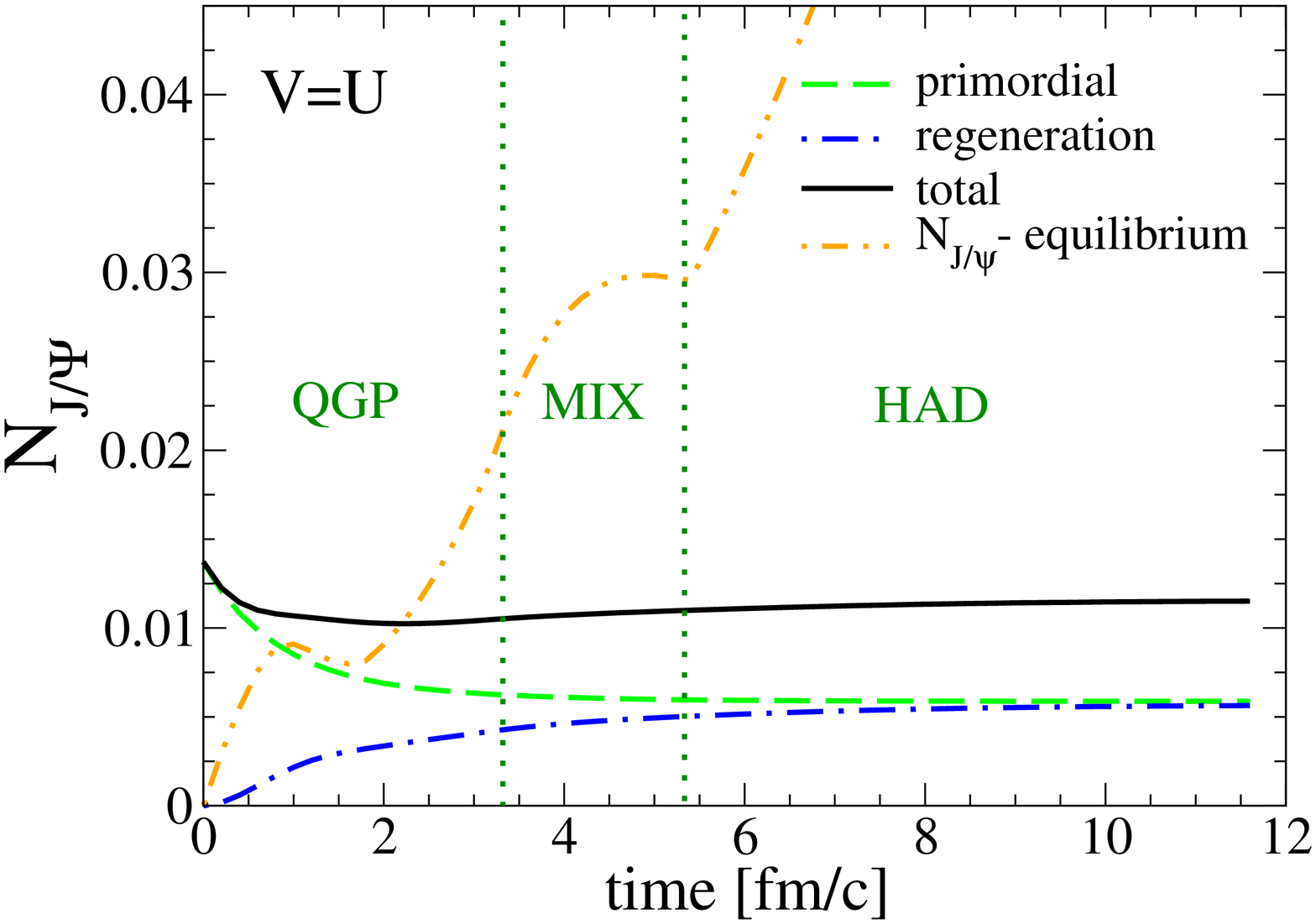}
                                                                         
\vspace{-0.8cm}
                                                                         
\includegraphics[angle=-90,width=0.48\textwidth]
{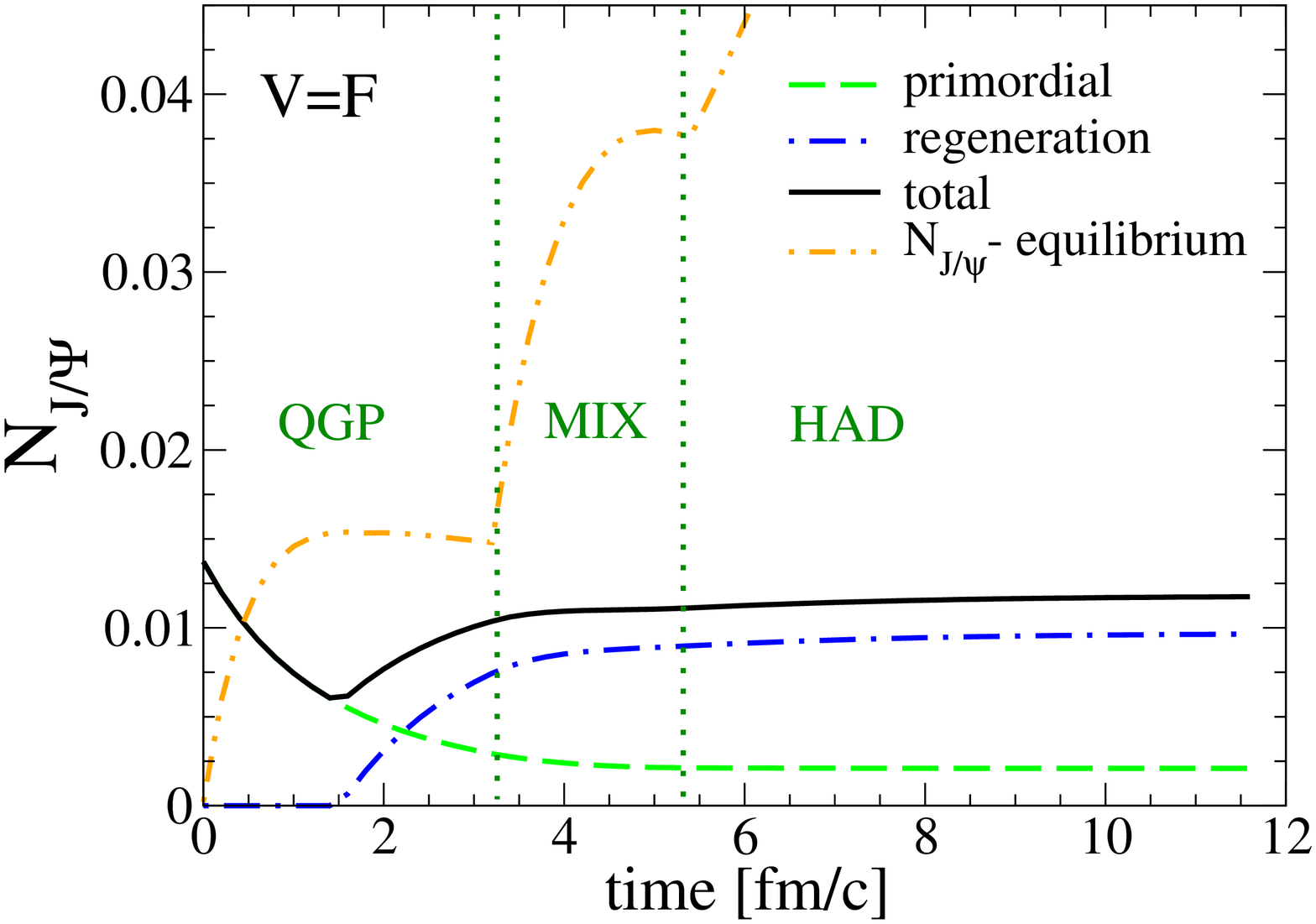}
\caption{(Color online) $J/\psi$ abundance as a function of time in 
  central ($N_{\rm part}$=380) Au-Au collisions at RHIC. Solid lines:
  total; dashed line: primordial component; dot-dashed line: regeneration
  component; double dot-dashed line: equilibrium limit of
  $J/\psi$, $N_{J/\psi}^{\rm eq}$. Upper panel: strong-binding
  scenario; lower panel: weak-binding scenario.}
\label{jpsi-time-evo}
\end{figure}
It is instructive to examine the time evolution of $J/\psi$ production
in the two scenarios, displayed in Fig.~\ref{jpsi-time-evo} for 
central collisions at mid rapidity at RHIC (excluding feeddown 
from $\chi_c$ and $\psi'$).
In both scenarios most of the dissociation and regeneration occur in
the QGP and the mixed phase, since the HG reaction rates are small. 
In the weak-binding scenario the time-dependent $J/\psi$ yield 
exhibits a ``dip'' structure around $\tau\simeq1.5$\,fm/$c$ because
the large dissociation rates suppress primordial $J/\psi$ very
rapidly and regeneration only starts after the medium temperature falls
below the $J/\psi$ dissociation temperature ($T_{J/\psi}^{\rm diss}
\simeq1.25\,T_c$). This scenario thus is closest in spirit to the 
statistical hadronization 
model~\cite{Andronic:2006ky} where all initial charmonia are suppressed 
(or never form to begin with, except for corona effects) and are then 
produced at the hadronization transition.

\subsection{Transverse-Momentum Spectra}
\label{ssec_pt-spec}

\begin{figure*}[!t]
\includegraphics[angle=-90,width=0.32\textwidth]
{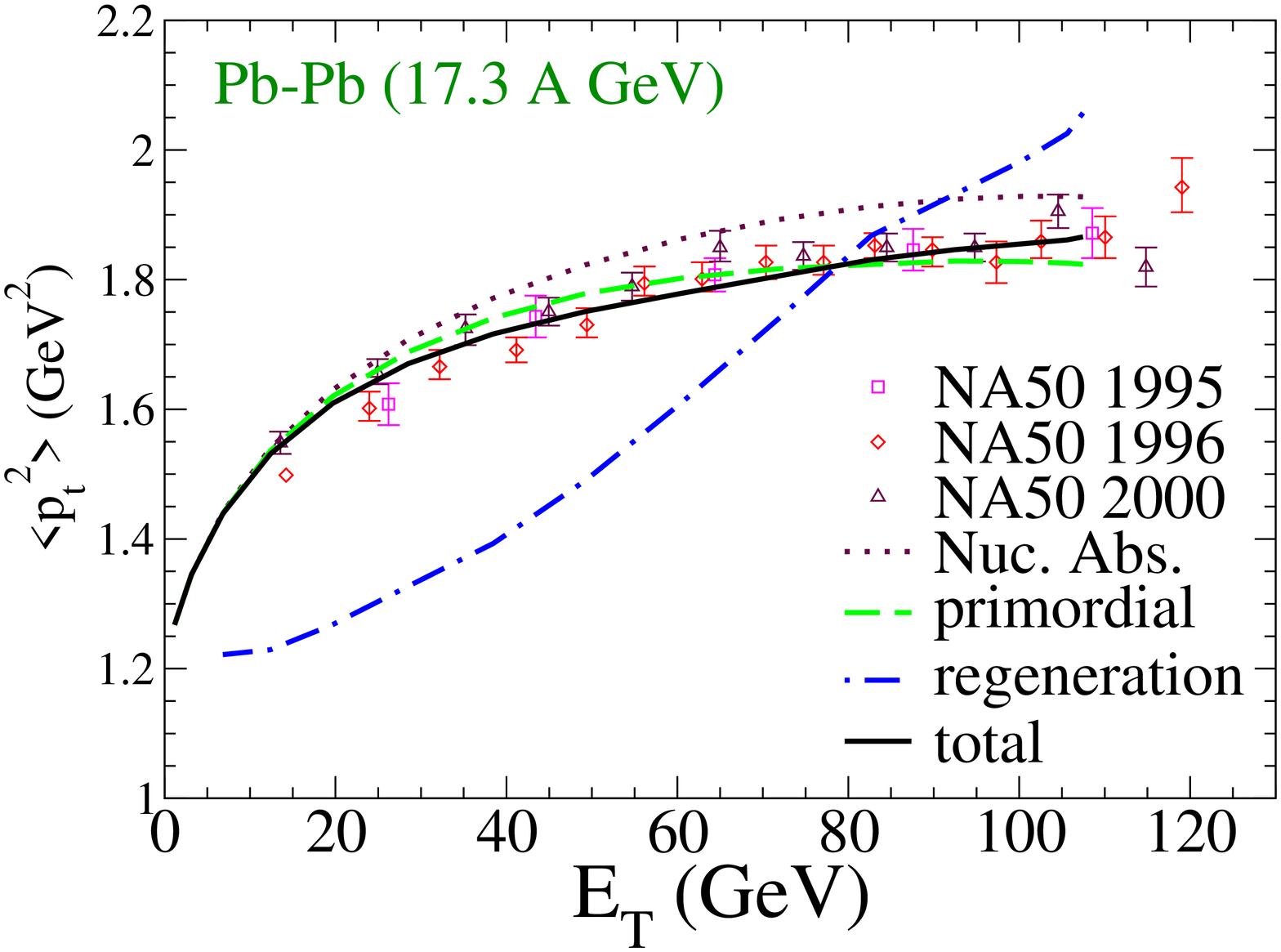}
\includegraphics[angle=-90,width=0.32\textwidth]
{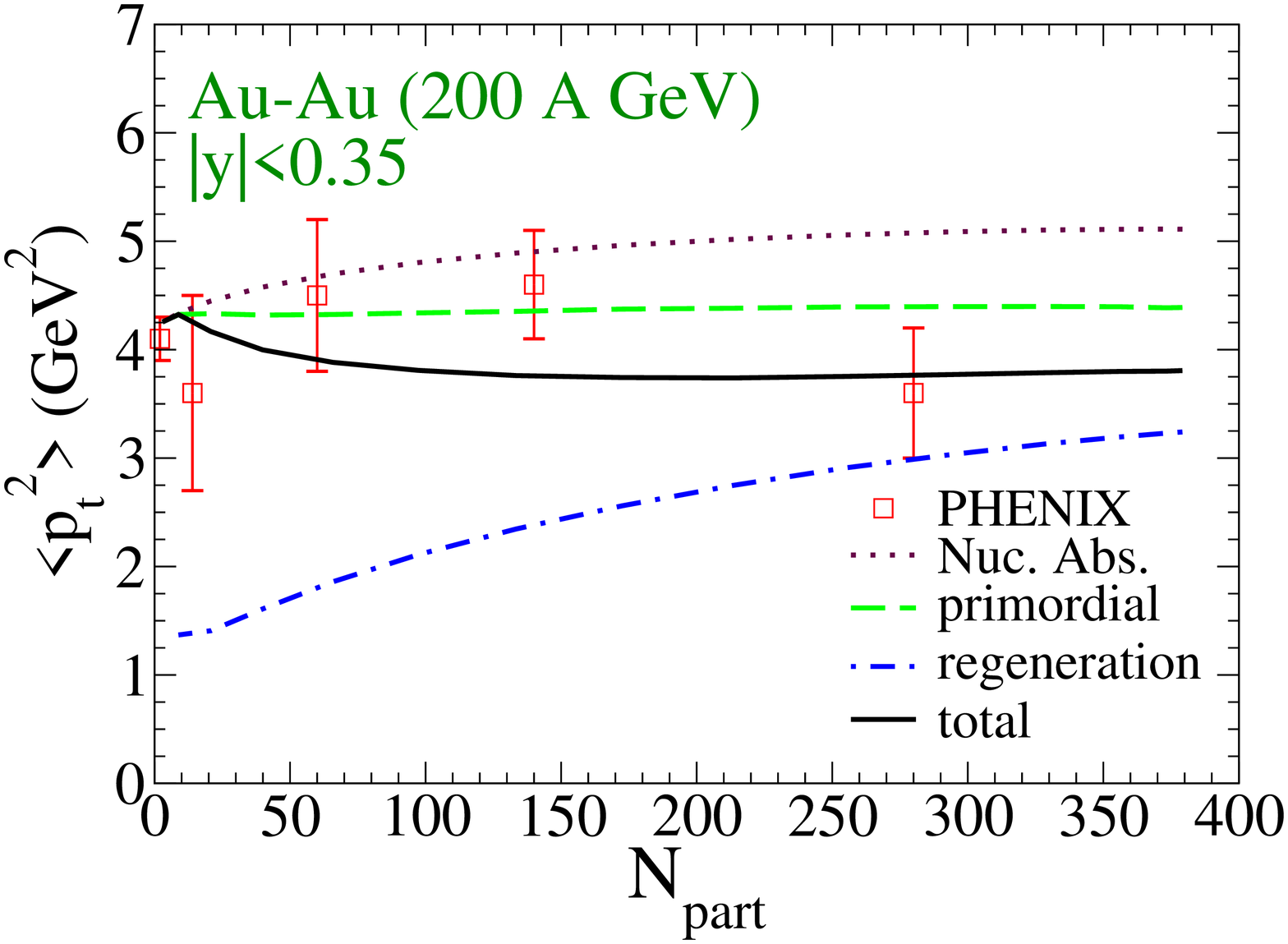}
\includegraphics[angle=-90,width=0.32\textwidth]
{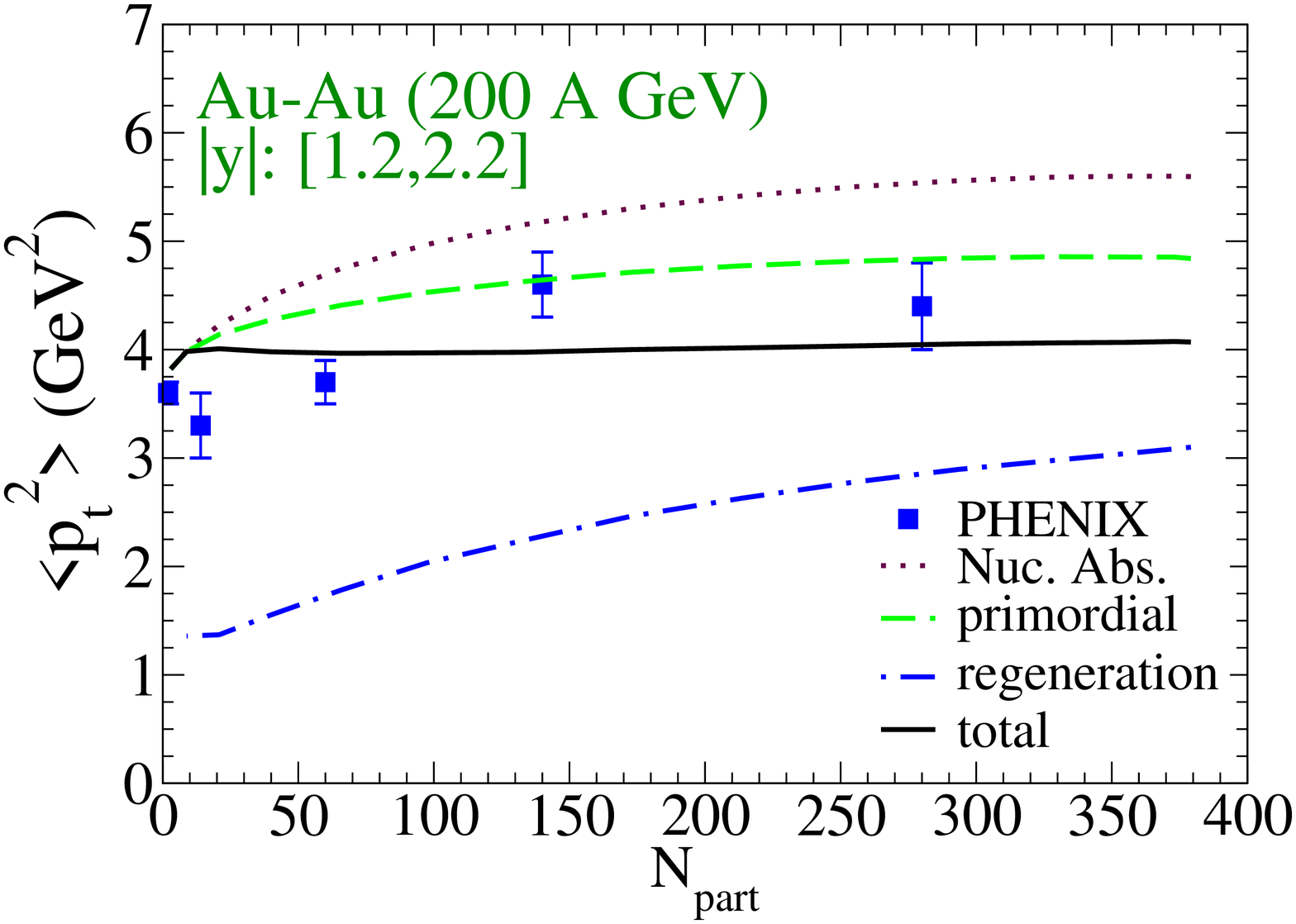}

\vspace{-0.5cm}

\includegraphics[angle=-90,width=0.32\textwidth]
{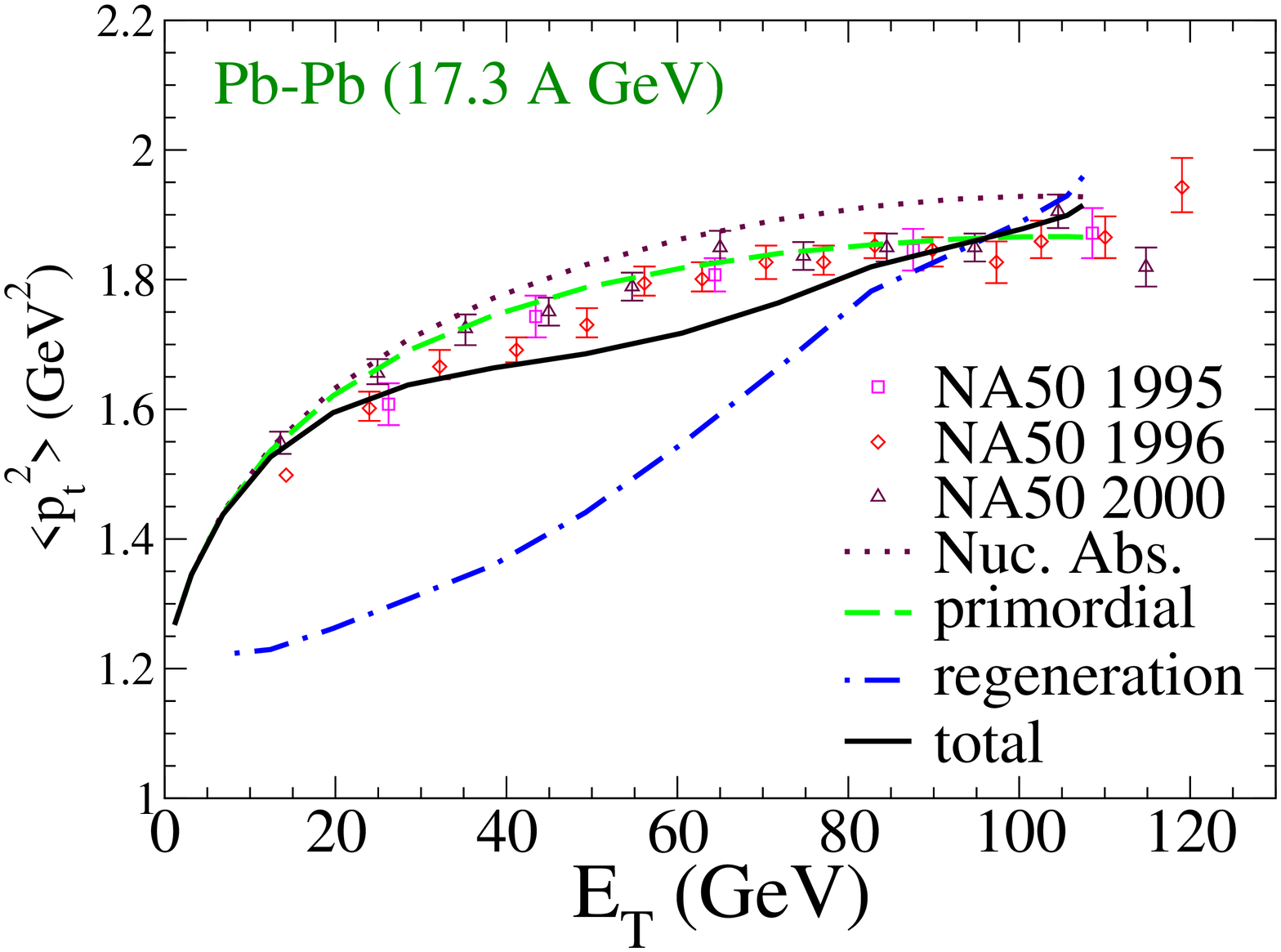}
\includegraphics[angle=-90,width=0.32\textwidth]
{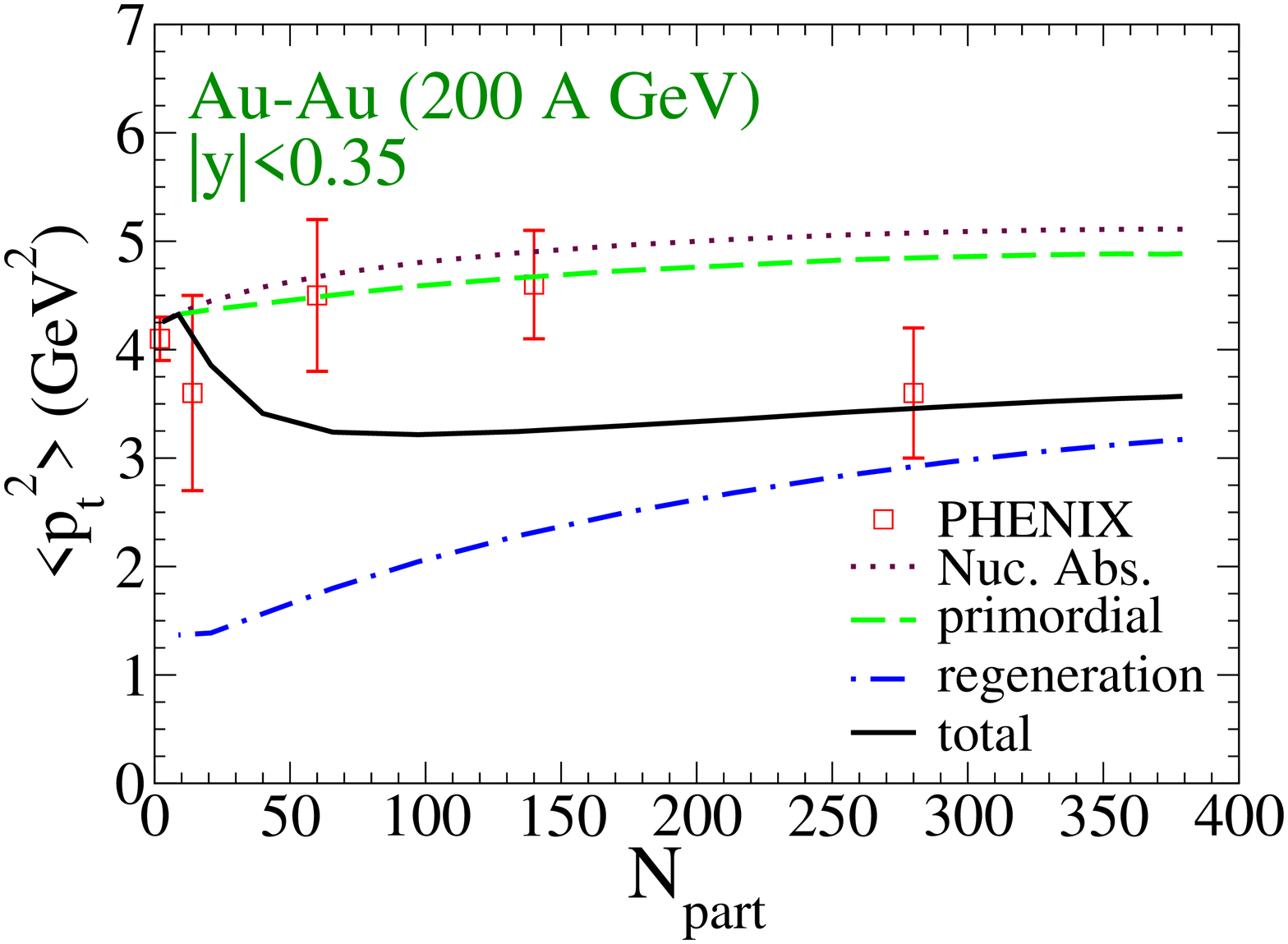}
\includegraphics[angle=-90,width=0.32\textwidth]
{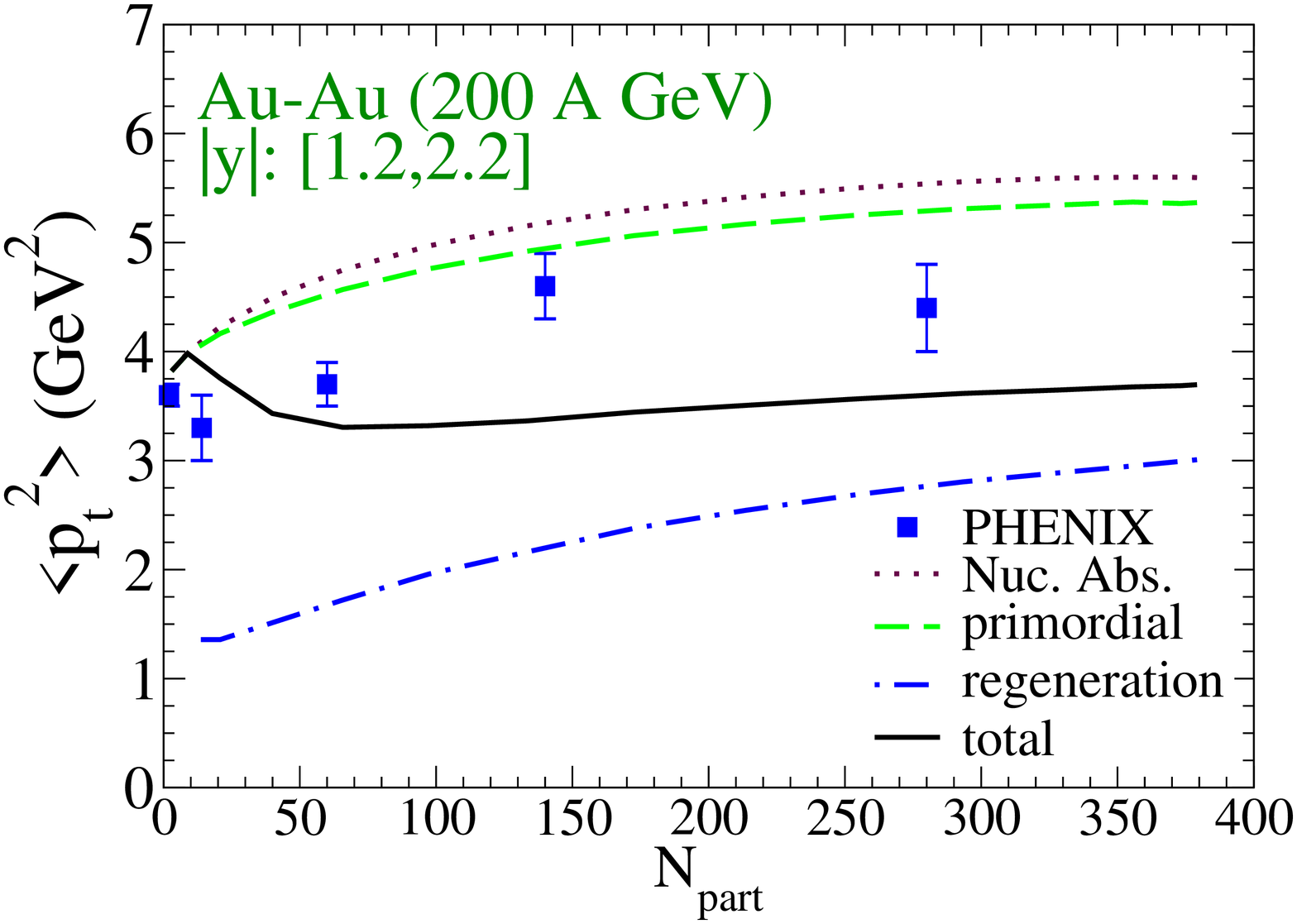}
\caption{(Color online) Results of the thermal rate-equation approach
 for $\langle p^2_t\rangle$ vs. centrality at SPS (left panels, compared 
 to NA50 data~\cite{Abreu:2000xe,Topilskaya:2003iy}) and RHIC at mid 
 and forward rapidity (middle and right panels, respectively, compared 
 to PHENIX data~\cite{Adare:2006ns}).
 In each panel, $\langle p^2_t\rangle$ is plotted for total $J/\psi$ 
 yield (primordial + regeneration component; solid lines), the
 suppressed primordial component (dashed line), the regeneration 
 component (dash-dotted line) and primordial production with CNM 
 effects only (dotted lines). 
 The upper (lower) panels correspond to the strong-binding 
 (weak-binding) scenario.}
\label{fig_pt2}
\end{figure*}

The results of the previous section suggest that, within the current
theoretical (e.g., charm-quark relaxation time $\tau_c^{\rm eq}$) and
experimental uncertainties both of the ``limiting" scenarios can
reproduce the centrality dependence of the inclusive
$R_{AA}^{J/\psi}(N_{\rm part})$ reasonably well at both SPS and RHIC
energies. However, the composition between suppression and regeneration
yields is rather different which ought to provide a key to
distinguish the two scenarios. The obvious ``lever arm"
are charmonium $p_t$ spectra~\cite{Grandchamp:2001pf}. One expects
that the primordial component is characterized by harder $p_t$ spectra
(following a power law at high $p_t$) while the regeneration component
produces softer $p_t$ spectra characterized by phase-space
overlap of (partially) thermalized charm-quark spectra.
However, in practice, the transition from the ``soft" recombination
regime to the ``hard" primordial regime is quite uncertain; \eg, 
collective flow and incomplete thermalization
of $c$-quarks can lead to a significant hardening of the regenerated
$J/\psi$ spectra, while a dissociation rate which increases with
3-momentum~\cite{Zhao:2007hh} can induce a softening of the spectra
of the surviving primordial charmonia.

For a concise discussion of the $p_t$ dependence of $J/\psi$
as a function of centrality at SPS and RHIC we here focus on the
average $p_t^2$, as compiled in Fig.~\ref{fig_pt2}.
At the SPS (left panels), the centrality dependence of 
$\langle p^2_t\rangle$ is largely dictated by the the Cronin effect 
in the primordial component, 
especially in the strong-binding scenario where this contribution 
dominates the yield at all centralities. The momentum dependence of the 
dissociation rate induces a slight suppression of $\langle p^2_t\rangle$ 
at large centrality compared to the case where only CNM effects are 
included (dashed vs. dotted line)~\cite{Zhao:2007hh}. In the 
weak-binding scenario, larger contributions from regeneration 
induce a slight ``dip" structure at intermediate centralities due to
a rather small collective flow at the end of the (relatively short)
mixed phase in these collisions. 

At RHIC energy (middle and right panels of Fig.~\ref{fig_pt2}), 
the individual primordial and regeneration components show qualitatively 
similar behavior for $\langle p^2_t\rangle(N_{\rm part})$ as at SPS,
\ie, an increase due to Cronin effect and collective flow,
respectively. At mid rapidity, the general trend is that with 
increasing centrality the growing regeneration contribution pulls 
down the average $\langle p^2_t\rangle$, in qualitative agreement
with the data. The curvature of the $\langle p^2_t\rangle(N_{\rm part})$
dependence, which appears to be negative in the data, is not
well reproduced, neither by the strong- nor by the weak-binding
scenario, even though the deviations are smaller in the former.
A more microscopic calculation of the gain term, together with more 
accurate estimates of the Cronin effect, are warranted to enable more 
definite conclusions. For both rapidity regions, the $\langle p^2_t\rangle$ of the suppressed primordial component is slightly larger in the weak- than in the strong-binding scenario. This is caused by the stronger 3-momentum increase of the dissociation rate in the strong-binding scenario, recall lower panel of Fig.~\ref{fig_diss-rate}.

The overall comparison to SPS and RHIC data for the $p_t$ dependence 
of $J/\psi$'s seems to indicate a slight preference for the 
strong-binding scenario. This is mostly derived from the observation
that for  peripheral collisions the experimentally observed 
$\langle p_t^2 \rangle$ essentially follows the extrapolation of the 
Cronin effect, suggesting $J/\psi$ production of predominantly
primordial origin (the collective flow imparted on the regeneration
component appears to be too small at these centralities). 

\begin{figure}[!t]
\centering
\includegraphics[angle=-90,width=0.48\textwidth]
{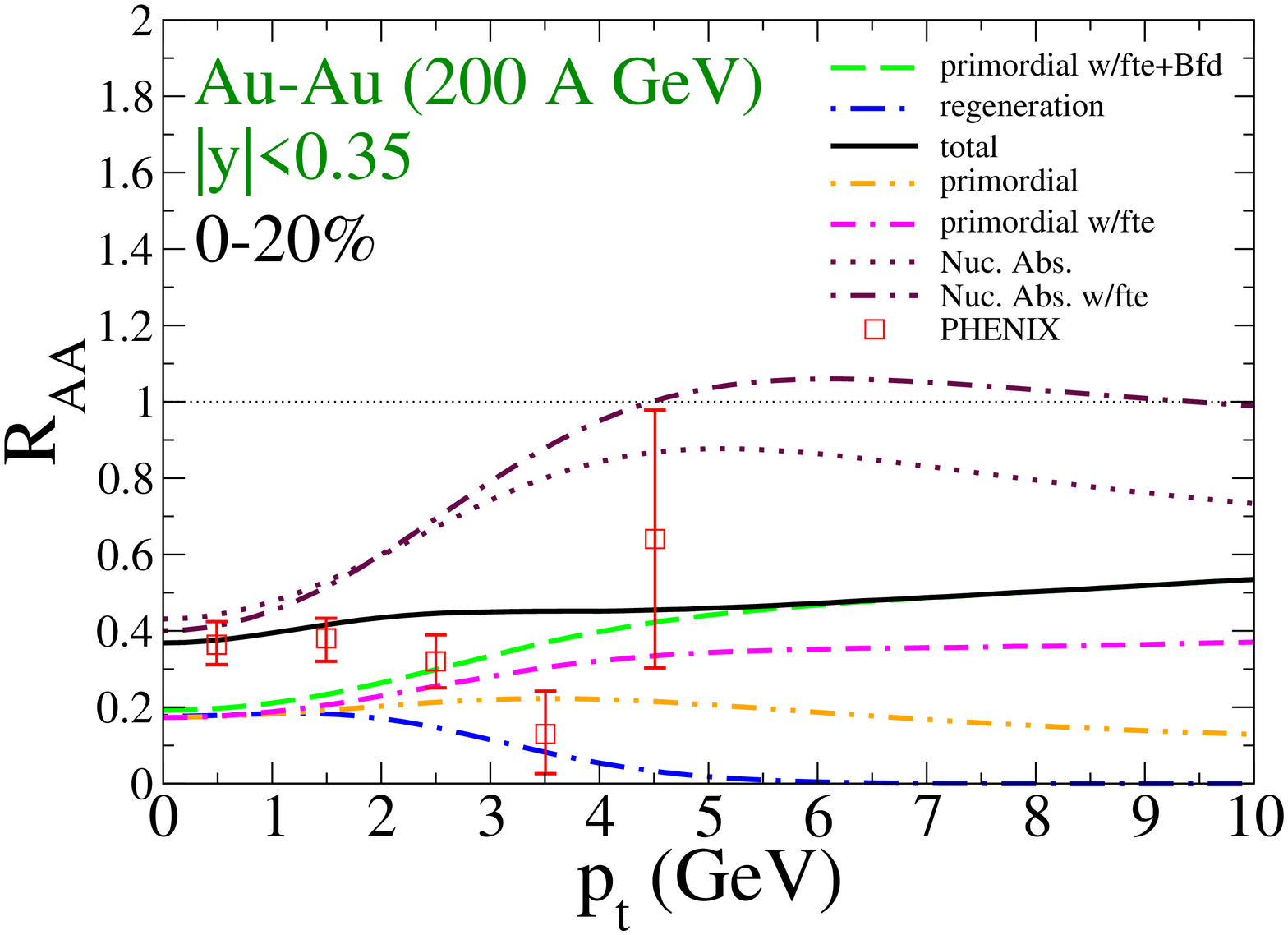}
                                                                          
\vspace{-0.65cm}
                                                                          
\includegraphics[angle=-90,width=0.48\textwidth]
{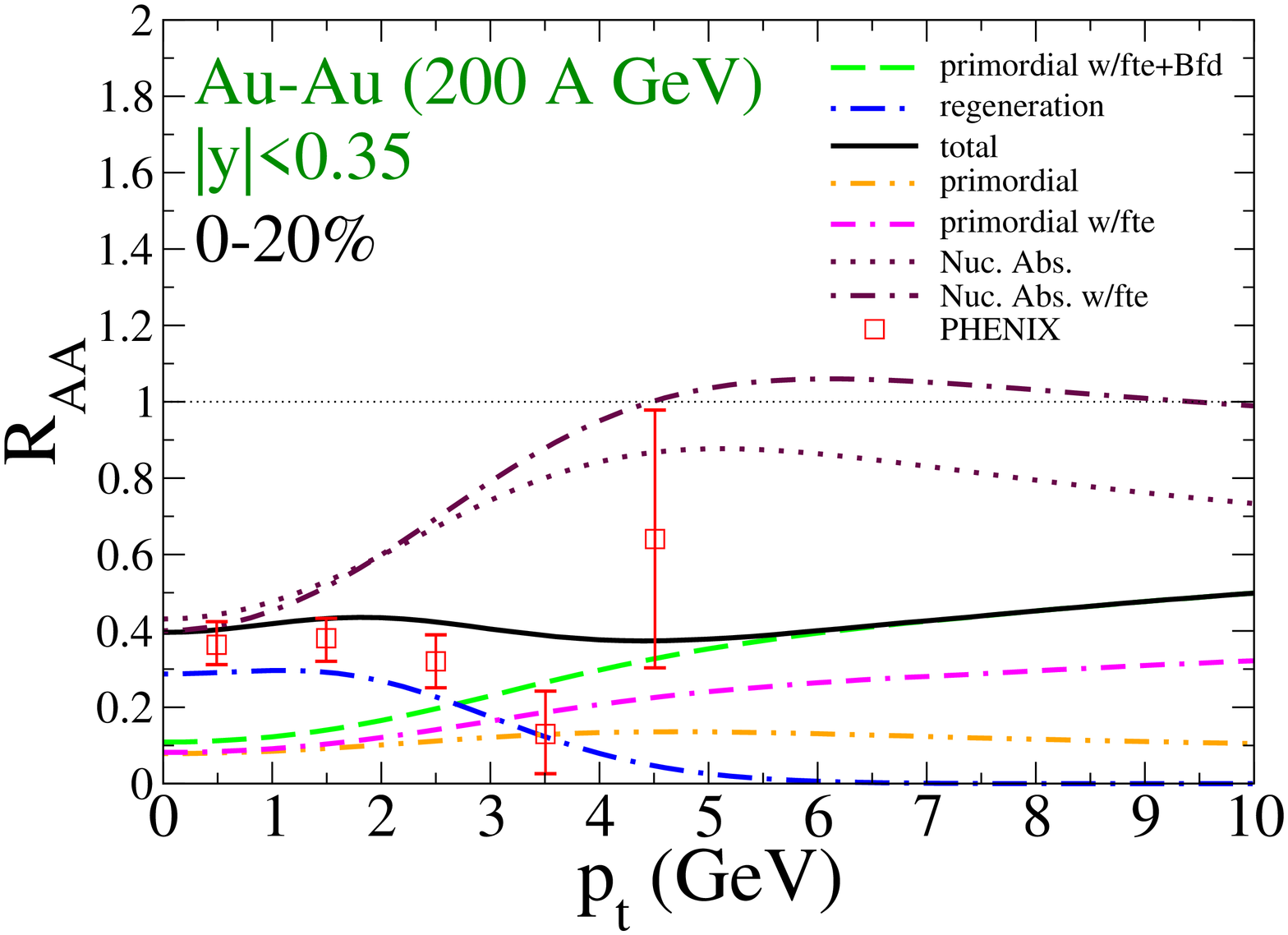}
\caption{Nuclear modification factor for $J/\psi$ $p_t$ spectra in 
central 200\,AGeV Au-Au collisions including formation-time effects 
(fte) and $B$-meson feeddown (Bfd) contributions. PHENIX 
data~\cite{Adare:2006ns} are compared to our rate-equation calculations 
in the strong- and weak-binding scenarios (upper and lower panel, 
respectively).}
\label{fig_raa-pt}
\end{figure}
The average $p_t^2$ mostly characterizes the momentum dependence of
charmonium production at low and moderate $p_t$ where most of the
yield is concentrated. Recent RHIC
data~\cite{Adare:2008sh,Abelev:2009qaa} have triggered considerable
interest in $J/\psi$ production at high $p_t\simeq5-10$\,GeV which is
expected to provide complementary information. It was found that the
suppression in $R_{AA}^{J/\psi}(p_t\gsim5\,{\rm GeV})$ in Cu-Cu
collisions is reduced compared to the low-$p_t$ region, with
$R_{AA}$-values of $\sim$0.7-1 or even
larger~\cite{Abelev:2009qaa}. This is quite surprising in light of the
light-hadron spectra measured thus far at RHIC which all exhibit
stronger suppression of $R_{AA}\simeq0.25$ for $p_t\gsim6$\,GeV (even
electron spectra from open heavy flavor, i.e., charm and bottom
decays). It also appears to be at variance with the thermal $J/\psi$
dissociation rates which, if anything, increase with momentum (recall
Fig.~\ref{fig_diss-rate}) and thus imply a stronger suppression at
higher $p_t$. Furthermore, the leakage effect referred to above is not
strong enough to produce the experimentally observed increase in
$R_{AA}^{J/\psi}(p_T\gsim5\,{\rm GeV})$~\cite{Zhao:2007hh}. However,
as has been pointed out more than 20 years
ago~\cite{Karsch:1988,Blaizot:1988ec}, the finite formation time of
the charmonium states, in connection with Lorentz time dilation at
high momentum, can lead to a reduction in the effective absorption
cross section (or thermal rate) as long as the bound-state wave
function has not developed (the pre-resonance state is more compact
than the fully formed $J/\psi$). Furthermore, for high-$p_t$ $J/\psi$
spectra significant feeddown contributions from the decay of
$B$-mesons~\cite{Xu:2008priv} is expected whose total yield is not
suppressed. In Ref.~\cite{Zhao:2008vu} we have schematically
implemented these effects into our rate-equation approach (as in the
present paper) and found that $R_{AA}^{J/\psi}$-values of up to
$\sim$0.85 at $p_t\simeq10$\,GeV~\cite{Zhao:2008vu} can be recovered
in Cu-Cu.  Here, we extend these calculations to provide predictions
for central Au-Au collision and specifically address the question
whether high-$p_t$ $J/\psi$ production can help to disentangle the
strong- and weak-binding scenarios. Here we do not readjust any
parameters relative to the above calculation which implies an
increase of 15-20\% in the total yield. Our results for
$R_{AA}^{J/\psi}(p_t)$ are displayed in Fig.~\ref{fig_raa-pt} up to
$p_t=10$\,GeV. We find that the suppression is reduced to about 0.5 at
the highest $p_t$, compared to about 0.4 at low $p_t$. This is similar
to the moderate enhancement found in the Cu-Cu case (where it went
from from $\sim$0.65 to 0.85).  More surprisingly, the high-$p_t$
suppression is very similar in both strong- and weak-binding
scenarios, despite the fact that the high-$p_t$ yield is exclusively
due to the primordial component whose strength is very different in
the 2 scenarios at low $p_t$. The reason is the 3-momentum dependence
of the dissociation rates, which become quite similar in the 2
scenarios at large 3-momentum: at $p\simeq 10$~GeV, the difference in
the energy-threshold due to binding energies of several 100\,MeV
becomes less relevant so that a collision with almost any thermal
parton is energetic enough for dissociating the bound state.

Let us finally comment on the elliptic flow, $v_2(p_t)$, of the 
$J/\psi$, which was hoped to be another good discriminator of
primordial and regenerated production. For the former, a nonzero
$v_2$ is basically due to the path-length difference when traversing the
azimuthally asymmetric fireball, typically not exceeding 2-3\%. For
the latter, much larger values can be obtained if the coalescing
charm quarks are close to thermalized. However, as pointed out in
Ref.~\cite{Zhao:2008vu}, the well-known mass effect suppresses the
$v_2(p_t)$ for heavy particles at $p_t\lsim m$; it is precisely in this
momentum regime where the regeneration component is prominent. Thus,
we predict that in both strong- and weak-binding scenarios the total
$J/\psi$ $v_2(p_t)$ does not exceed $\sim$3\% at any $p_t$. The only
alternative option we can envision are strong elastic interactions of 
the $J/\psi$ which is only conceivable in the strong binding scenario to 
avoid break-up in scattering off thermal partons~\cite{Rapp:2009my}.

\section{Conclusion}
\label{sec_concl}
The main goal of this work was to construct charmonium spectral 
functions which are constrained by thermal lattice-QCD computations,
and to apply them to experimental data in heavy-ion collisions. 
Employing a thermal rate equation we have implemented equilibrium 
properties of charmonia (binding energy, constituent charm-quark mass 
and dissociation rate) as extracted from a thermodynamic $T$-matrix 
calculation. The resulting spectral functions for two ``limiting" 
scenarios, with small and large $J/\psi$ dissociation temperatures, 
have been verified to produce a weak temperature dependence in pertinent 
euclidean correlators, roughly compatible with lQCD.  
We have argued that these two scenarios may serve as generic 
representatives for charmonium kinetics in heavy-ion 
collisions, bracketing strong and weak in-medium binding. Therefore, 
we believe that the qualitative conclusions drawn from these two scenarios 
should be rather model-independent. Within current theoretical uncertainties
(especially for the degree of charm-quark relaxation and its impact
on the $J/\psi$ regeneration yield) and essentially two fit parameters
($\tau_c$ and $\alpha_s$ controlling the regeneration yield and 
suppression strength) both scenarios can reproduce SPS and RHIC data 
for the centrality dependence of inclusive $J/\psi$ production
reasonably well. However, the partition of primordial and regenerated
yields is quite different in the two scenarios: the former dominates
for strong binding (down to 50\% in central Au-Au at RHIC), while for 
weak binding regeneration largely prevails at RHIC energies (except
for peripheral collisions). We have investigated the $p_t$ dependence
of the $J/\psi$ yield in both scenarios and found that differences in
the average $p_t^2$ reach up to 20\% in semicentral Au-Au collisions. 
The strong-binding scenario seems slightly favored in this observable, 
but theoretical uncertainties (e.g., the blast-wave treatment of the
regeneration component) prevent us from definite conclusions at this 
point. We also have to await more accurate $A$-$A$ and $p$(d)-$A$ data,
where the latter determine the input in terms of shadowing, 
nuclear absorption and Cronin effects.

Further developments of the theoretical approach are in order, and we 
plan to improve our calculations in several respects. First, as mentioned 
before, work is in progress employing a Boltzmann transport equation 
(\ref{boltz}) extended by a microscopic treatment of the gain term to 
treat charmonia regeneration and dissociation on the same footing. This 
will enable an explicit account of (time-dependent) charm-quark phase-space
distribution in $c$-$\bar c$ recombination reactions, as following, 
\eg, from realistic Langevin simulations~\cite{vanHees:2007me} with 
constraints from the $T$-matrix formalism and from open-charm observables.
Second, hydrodynamic simulation of the medium evolution could be 
employed for a more detailed and realistic description of 
the temperature and the flow field of the underlying medium, 
especially in coordinate space.
Third, a microscopic model for primordial $c\bar c$ and charmonium
production is warranted to better disentangle nuclear shadowing and 
absorption in the pre-equilibrium stage, including formation-time
effects. This would improve the initial conditions to the kinetic 
approach in the hot medium. 
These developments will eventually produce a comprehensive approach 
which can serve as a quantitative bridge between theoretical studies 
of charmonia in the QGP (and HG) and their phenomenology in heavy-ion 
collisions. As our investigations in the present paper have shown, 
quantitative studies at the 10-20\% level are needed to deduce basic 
properties of the strong force (such as color screening of Coulomb and 
confining interactions) from heavy-ion collisions. These insights are
also pivotal to improve our knowledge of the phase structure of hot and 
dense QCD matter.

\acknowledgments We are grateful to L.~Grandchamp for providing us
with his codes. We thank A.~Frawley for useful discussions.  This work
is supported by a US National Science Foundation CAREER award under
grant No. PHY-0449489, by NSF grant PHY-0969394 and by the
A.-v.-Humboldt foundation (Germany).

\end{document}